\documentclass[aps,prc,showpacs]{revtex4-1}
\usepackage{amsmath} 
\usepackage{amssymb} 
\usepackage{graphics}
\usepackage{epsfig}
\usepackage{slashed}
\newcommand\nn{\nonumber}
\newcommand\ba{\begin{eqnarray}}
\newcommand\ea{\end{eqnarray}}
\newcommand\alb{\begin{align}}
\newcommand\ale{\end{align}}
\newcommand\be{\begin{equation}}
\newcommand\ee{\end{equation}}

\usepackage{longtable}
\usepackage{latexsym}

\usepackage{amssymb}
\usepackage{amsfonts} 
\usepackage{amsmath}

\begin{document}
\title{Radiative corrections to elastic proton-electron scattering measured in coincidence}

\author{G.I.~Gakh}

\affiliation{\it NSC ''Kharkov Institute of Physics and Technology'',
Akademicheskaya, 1, 61108 Kharkov,  and \\
V.N.~Karazin Kharkiv National University, 61022 Kharkov, Ukraine}

\author{M.I.~Konchatnij}
\affiliation{\it NSC ''Kharkov Institute of Physics and Technology'',
Akademicheskaya, 1, 61108 Kharkov, and \\
V.N.~Karazin Kharkiv National University, 61022 Kharkov, Ukraine}

\author{N.P.~ Merenkov}
\affiliation{\it NSC ''Kharkov Institute of Physics and Technology'',
Akademicheskaya, 1, 61108 Kharkov,  and \\
 V.N.~Karazin Kharkiv National University, 61022 Kharkov, Ukraine}

\author{Egle~Tomasi-Gustafsson}
\affiliation{\it IRFU, CEA, Universit\'e Paris-Saclay, 91191 Gif-sur-Yvette Cedex, France}

\date{\today}

\begin{abstract}
The differential cross section for  elastic scattering of
protons on electrons at rest is calculated taking into account the QED radiative
corrections to the leptonic part of interaction. These model-independent radiative corrections
arise due to emission of the virtual and real soft and hard photons as well as to vacuum polarization.
We analyze an experimental setup when both the final particles are recorded in coincidence and their energies are determined within some uncertainties. The kinematics, the cross section, and the radiative corrections are calculated and numerical results are presented.
\end{abstract}

\maketitle

\section{Introduction}

The polarized and unpolarized scattering of electrons off protons has
been widely studied, as it is considered the simpler way to access
information on the proton structure, assuming that the interaction occurs through the exchange of a virtual photon of four-momentum $q$. The experimental
determination of the elastic proton electromagnetic form factors in
the region of small and large momentum transfer squared is one
of the major field of research in hadron physics (see the review \cite{Pacetti:2015iqa}). New experimental
possibilities allowed to reach better precision and to perform
polarization experiments as earlier suggested in Refs. \cite{Akhiezer:1968ek,Akhiezer:1974em}.

The determination of the proton electromagnetic form factors, at
$Q^2=-q^2\geq 1$ GeV$^2$, from polarization observables
showed a surprising result: the polarized and unpolarized
experiments, although based on the same theoretical background (same
formalism and same assumptions), ended up with inconsistent values
of the form factor ratio (see \cite{Puckett:2010ac} and references therein). Possible explanation of this
discrepancy is to take into account higher order radiative
corrections \cite{Kuraev:2013dra,Gramolin:2016hjt} including the interference between one and two photon
exchange \cite{Arrington:2003ck}, correlations among parameters  \cite{TomasiGustafsson:2006pa}, normalization  of data \cite{Pacetti:2016tqi}. This puzzle has given rise to many speculations and
different interpretations, suggesting further experiments (for a
review, see Ref. \cite{Perdrisat:2006hj}).

In the region of small $Q^2$ one can determine the proton charge
radius ($r_E$) which is one of the fundamental quantities in
physics. Precise knowledge of its value is important for the
understanding of the structure of the nucleon in the theory of
strong interactions (QCD) as well as in the spectroscopy of atomic hydrogen.

Recently, the determination of the $r_E$ with
muonic atoms lead to the so-called proton radius puzzle.
Experiments on muonic hydrogen by laser
spectroscopy measurement of the $\mu $p (2S-2P) transition frequency,
in particular the latest result on the proton charge radius
\cite{Pohl:2010zza,Antognini:1900ns} $r_E=0.84087(39)$ fm,  is one
order of magnitude more precise but smaller by seven standard
deviations compared to the average value $r_E=0.8775(51)$ fm which
is recommended by the 2010-CODATA review \cite{Mohr:2015ccw}. The CODATA value is
obtained coherently from hydrogen atom spectroscopy and electron-proton elastic
scattering measurements. The latest experiments with electrons at
Jlab \cite{Zhan:2011ji} and MAMI \cite{Bernauer:2010wm} confirm this value, and,  therefore, do
not agree with the results on the proton radius from of the laser spectroscopy of the muonic
hydrogen.

While the corrections to the laser spectroscopy experiments seem
well under control in the frame of QED and may be estimated with a
precision better than 0.1\%, in case of electron-proton elastic
scattering the best achieved precision is of the
order of few percent. Different sources of possible systematic
errors of the muonic experiment have been discussed, however no
definite explanation of this difference has been given yet (see Ref.
\cite{Antognini:2011zza} and references therein).

The proton radius puzzle lead to a large number of theoretical papers suggesting
solutions based on different approaches, as new physics beyond
the Standard Model \cite{TuckerSmith:2010ra,Batell:2011qq}. Other approaches analyze the extraction of
the proton radius from the electron-proton scattering
data. In Ref. \cite{Giannini:2013bra}, it is argued that a proper Lorentz
transformation of the form factors accounts for the
discrepancy. The authors of Ref. \cite{Kraus:2014qua} stated that radius
extraction with Taylor series expansions cannot be trusted to be
reliable. A fit function based on a conformal mapping was used in
Ref. \cite{Lorenz:2012tm,Lorenz:2014yda}. The extracted value of the proton radius agrees
with the one obtained from muonic hydrogen. A similar result was
obtained in Ref.\cite{Horbatsch:2016ilr} using the approach based on the chiral perturbation
theory \cite{Peset:2014jxa}.

In Ref. \cite{Griffioen:2015hta}, the
authors argued that the proton radius puzzle can be explained by
truncating the electron scattering data to low momentum transfer.
But the authors of Ref. \cite{Distler:2015rkm} showed that the procedure is
inconsistent and violates the Fourier theorem. The authors of the
paper \cite{Bernauer:2016ziz} inspected several recent refits of the Mainz data, that
result in small radii and found flaws of various kinds in
all of them.

A recent review summarizes the current state of the problem and
gives an overview over upcoming experiments \cite{Bernauer:2014nqa}.

More experiments   in the region of small
$Q^2$ are expected to shed some light on this intriguing problem.
The PRad collaboration \cite{Gasparian:2014rna} is currently preparing a
new magnetic-spectrometer-free electron-proton scattering experiment
in Hall B at Jefferson Lab for a new independent measurement of
$r_E$. This will allow to reach extremely low $Q^2$ range (10$^{-4}$
- 10$^{-2}$) (GeV/c)$^2$ with an incident electron beams with energy of few GeV. The
lowest $Q^2$ range measured up to date is in the recent Mainz
experiment \cite{Bernauer:2010wm} where the minimum value of $Q^2$ was
3 $\times$ 10$^{-3}$ (GeV/c)$^2$. Reaching low $Q^2$ range is
critically important since the charge radius of the proton is
extracted as the slope of the measured electric form factor $G_p^
E(Q^2)$ for $Q^2\to 0$, requiring an extrapolation. The MUSE
experiment \cite{Downie:2016mry} (PSI, Switzerland)
will simultaneously measure elastic electron and muon
scattering on the proton, in both charge states. The expected precision on
cross section measurements for the elastic scattering of $\mu^{+/-}$
and e$^{+/-}$ is better than the percent,  over a $Q^2$ range
from 0.002 to 0.07 GeV$^2$.
Low energy $ep$ scattering experiments are also planned at the PRAE platform \cite{Voutier:2016}, making use of a high intensity low energy  electron beam and with a very precise measurement of the electron angle and energy. At the Mainz Microtron the simultaneous detection of the proton and the electron is proposed \cite{Vorobyev:2016}, in the measurement  of the absolute cross section at  per mille absolute precision.

Recently, we suggested that proton elastic scattering on atomic
electrons may allow a precise measurement of the proton charge radius
\cite{Gakh:2013pda}. The main advantage of this proposal is that inverse
kinematics allows one to access  very
small values of the transferred momenta, up to four orders of
magnitude smaller than the ones presently achieved, where the cross section is huge.
Moreover, the applied radiative corrections differ essentially, as the electron mass should be taken explicitly into account.
The unpolarized and polarized observables for the elastic scattering of a proton projectile
on an electron target were derived in Ref.  \cite{Gakh:2011sh} and references therein.  Although we are aware that an experiment measuring the elastic cross
section at very small $Q^2$ can not, by itself, produce a constraint on the slope of form factors,
and therefore a precise extraction of the radius, a combined series of low $Q^2$ very precise
measurements, combined with a physical parametrization of form factors, will help for a meaningful
extrapolation to the static point.

The inverse kinematics was previously used in a number of the
experiments to measure the pion or kaon radius from the elastic
scattering of negative pions (kaons) from electrons in a
liquid-hydrogen target. The first experiment was done at Serpukhov
\cite{Adylov:1974sg} with pion beam energy of 50 GeV. Later, a few
experiments were done at Fermilab with pion beam energy of 100 GeV
\cite{Dally:1977vt} and 250 GeV \cite{Dally:1982zk}. At this laboratory, the
electromagnetic form factors of negative kaons were measured
by direct scattering of 250 GeV kaons from stationary electrons
\cite{Dally:1980dj}. The typical values of the radiative corrections in this
case are of the order of 7-10\% \cite{Kahane:1964zz,Bardin:1970yn}. Later on, a measurement of
the pion electromagnetic form factor was done at the CERN SPS
\cite{Amendolia:1986wj,Amendolia:1984nz} by measuring the interaction of 300 GeV pions with
the electrons in a liquid hydrogen target. This experiment  measured
only the angles of the final particles to select the pion-electron elastic events,
whereas, in previous experiments, both 3-momenta were measured.

The use of the inverse kinematics is proposed in a new experiment at
CERN \cite{Abbiendi:2016xup} to measure the running of the
fine-structure constant in the space-like region by scattering
high-energy muons (with energy 150 GeV) on atomic electrons,
$\mu e\to \mu e$. The aim of the experiment is the extraction of the hadron vacuum polarization contribution.
The proposed technique will be similar to the
one described in \cite{Amendolia:1986wj,Amendolia:1984nz} for the measurement of the pion form
factor: a precise measurement of the scattering angles of both
outgoing particles.

For the analysis of the results of the possible experiment on
the elastic proton-electron scattering, it is necessary to take into
account radiative corrections. In this paper we calculate
the model-independent QED radiative corrections to the differential
cross section of the elastic scattering of the protons on
electrons at rest. The radiative corrections due to the emission of
virtual and real (soft and hard) photons in the electron vertex as
well the vacuum polarization are taken into account. The corresponding Feynmann diagrams are shown in Fig.\,1.
We consider an
experimental setup where the final particles are detected in
coincidence and their energies are measures within some uncertainty.
Numerical estimations of these corrections in considered case are given and their dependence on the kinematical variables is illustrated.

\begin{figure}[t]
\centering
\includegraphics[width=0.44\textwidth]{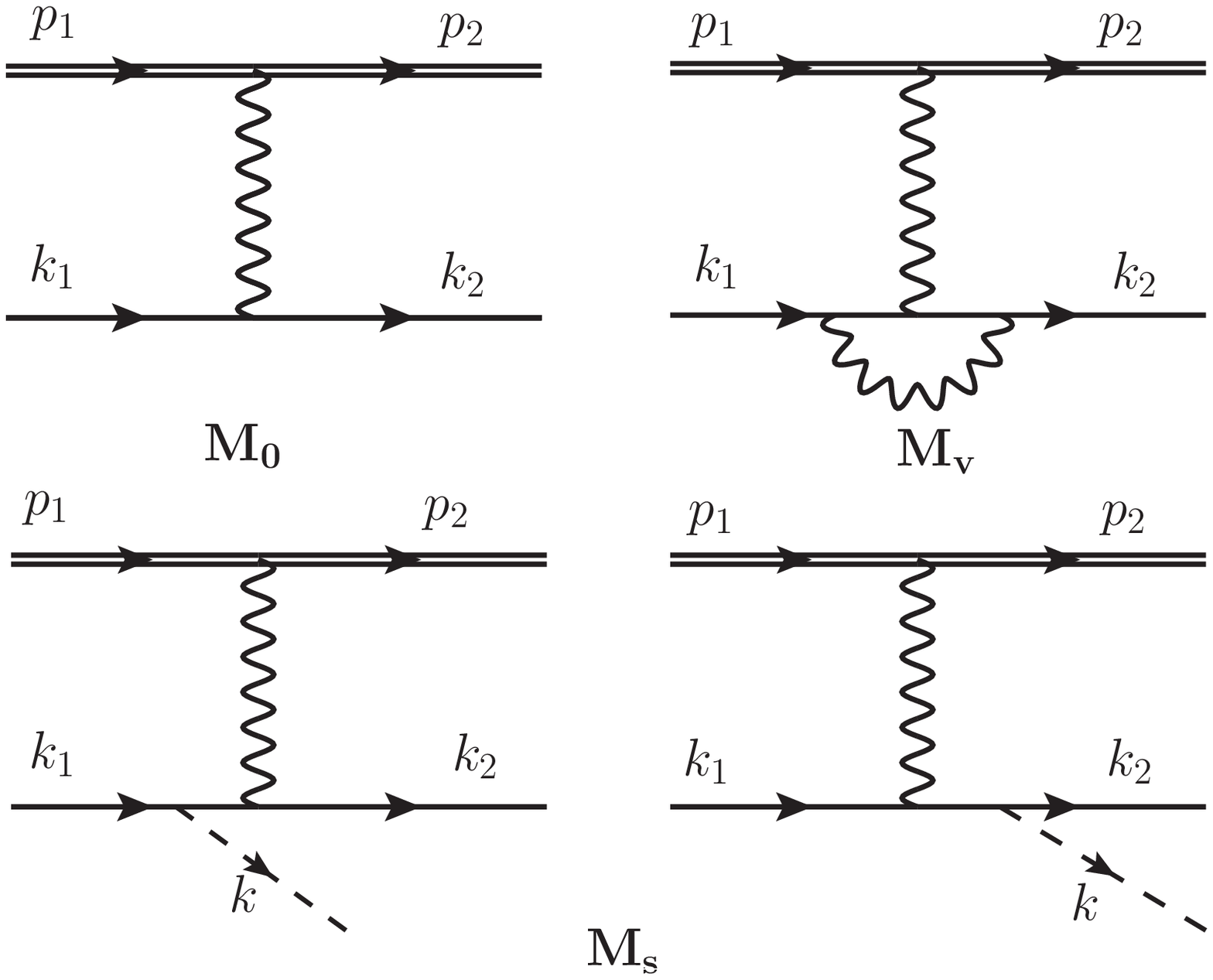}
\includegraphics[width=0.46\textwidth]{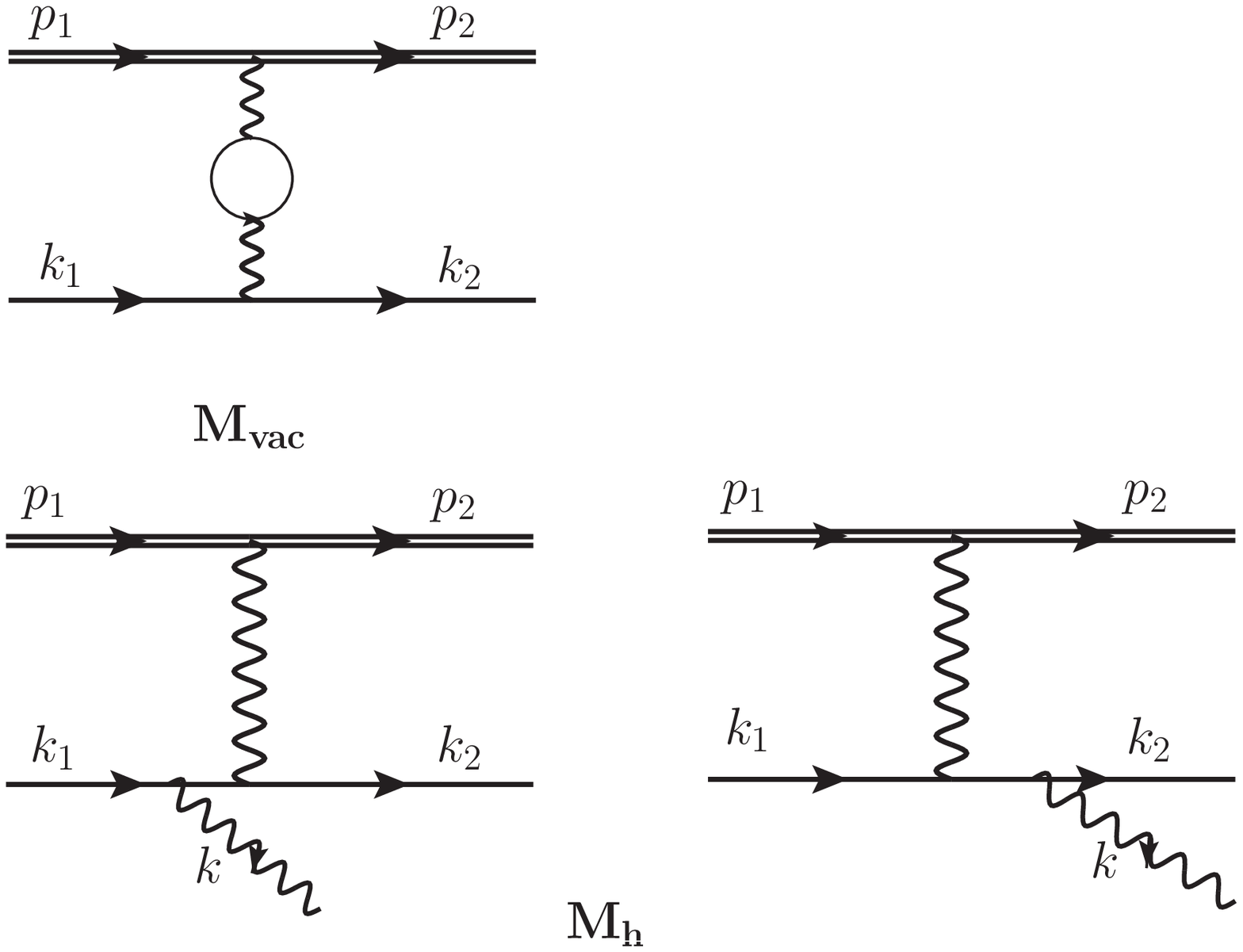}
\caption{Feynman's diagrams corresponding to the Born
 approximation and first order virtual  radiative corrections (top), and to initial and final real photon emission,
  $M_s$ (soft) and $M_h$ (hard), from the lepton vertex (bottom). }
 \label{diagrams}
\end{figure}

\section{Formalism}
Let us consider the reaction
\be
p (p_1) + e(k_1) \to p(p_2)+e(k_2),
\label{eq:1}
\ee
where the particle momenta are indicated in parenthesis, and
$q=k_1-k_2=p_2-p_1$ is the four momentum of the virtual photon.

\subsection{Inverse kinematics}

A general characteristic of all reactions of elastic and inelastic
hadron scattering by atomic electrons (which can be considered at
rest) is the small value of the  momentum transfer squared, even for
relatively large energies of the colliding particles. Let us give
details of the order of magnitude and the dependence of the
kinematic variables, as they are very specific for these reactions.
In particular, the electron mass can not be neglected in the
kinematics and dynamics of the reaction, even when the beam energy is of the order of  GeV.

One can show that, for a given energy of the proton beam, the
maximum value of the four momentum transfer squared, in the
scattering on electrons at rest, is
\be
\label{eq:2}
(Q^2)_{max}=\frac{4m^2\vec p^2}{M^2+2mE+m^2},
\ee
where m(M) is the electron (proton) mass, $E (\vec{p})$
is the energy (momentum) of the proton beam. Being proportional to
the electron mass squared, the four momentum transfer squared is
restricted to very small values, where the proton can be considered
structureless.

The four momentum transfer squared is expressed as a function of the
energy of the scattered electron, $\epsilon_2$, as:
$q^2=(k_1-k_2)^2=2m(m-\epsilon_2)$, where
\be\label{eq:3}
\epsilon_2=m\frac{(E+m)^2+\vec{p}^2\cos^2\theta_e}{(E+m)^2-\vec{p}^2\cos^2\theta_e},
\ee
where $\theta_e$ is the angle between the proton beam and the
scattered electron momenta.

From energy and momentum conservation, one finds the following
relation between the angle and the energy of the scattered electron:
\be\label{eq:4}
\cos\theta_e=\displaystyle\frac{(E+m)(\epsilon_2-m)} {|\vec
p|\sqrt{\epsilon_2^2-m^2}},
\ee
which shows that $\cos\theta_e\geq 0$ (the electron can never be
scattered backward). One can see from Eq. (\ref{eq:3}) that, in  inverse
kinematics, the available kinematical region is reduced to small
values of $\epsilon_2$:
\be\label{eq:5}
\epsilon_{2,max}=m\frac{2E(E+m)+m^2-M^2}{M^2+2mE+m^2},
\ee
which is proportional to the electron mass. From the momentum
conservation, on can find the following relation between the energy
and the angle of the scattered proton $E_2$ and $\theta_p$:
\be\label{eq:6}
E_2^{\pm}=\frac {(E+m)(M^2+mE)\pm M\vec p^2
\cos\theta_p\sqrt{\frac{m^2}{M^2} -\sin^2\theta_p}}{(E+m)^2-\vec p^2
\cos^2\theta_p},
\ee
showing that, for one proton angle, there may be two values of
the proton energies, (and two corresponding values for the
recoil-electron energy and angle as well as for the transferred
momentum $q^2$). This is a typical situation when the center-of-mass
velocity is larger than the velocity of the projectile in the center
of mass, where all the angles are allowed for the recoil electron.
The two solutions coincide when the angle between the initial and
final hadron takes its maximum value, which is determined by the
ratio of the electron and scattered hadron masses $M_h$,
$\sin\theta_{h,max}=m/M_h$. One concludes that hadrons are scattered
on atomic electrons at very small angles, and that the larger is
the hadron mass, the smaller is the available angular range for the
scattered hadron.

\subsection{Differential cross section}

In the one-photon exchange (Born) approximation, the matrix element
${\cal M}^{(B)}$ of the  reaction (1) can be written as:
\be\label{eq:7}
{\cal M}^{(B)}=\frac{e^2}{q^2}j_{\mu}J_{\mu},
\ee
where $j_{\mu}(J_{\mu})$ is the leptonic (hadronic) electromagnetic
current:
\ba
j_{\mu}&=&\bar u(k_2)\gamma_{\mu} u(k_1),
\label{eq:8}\\
J_{\mu}&=&\bar u(p_2)\left [F_1(q^2)\gamma_{\mu}- \frac{1}{2M}
F_2(q^2)\sigma_{\mu\nu}q_{\nu}\right ] u(p_1)=\nn\\
&=&\bar u(p_2)\left [G_M(q^2)\gamma_{\mu}- F_2(q^2)P_{\mu}\right ]
u(p_1),
\nn
\ea
 where $P_{\mu}=(p_1+p_2)_{\mu}/(2M)$. $F_1(q^2)$ and $F_2(q^2)$ are the Dirac and Pauli
proton electromagnetic form factors, $G_M(q^2)=F_1(q^2)+F_2(q^2)$ is
the Sachs proton magnetic form factor. The matrix element squared is written as:
\be\label{eq:9}
|{\cal
M}^{(B)}|^2=16\pi^2\frac{\alpha^2}{q^4}L_{\mu\nu}W_{\mu\nu},\mbox{~with~}
L_{\mu\nu}=j_{\mu}j_{\nu}^*,~W_{\mu\nu}=J_{\mu}J_{\nu}^*,
\ee
where $\alpha=e^2/(4\pi)=1/137$ is the electromagnetic fine structure constant.
The leptonic tensor, $L_{\mu\nu}$, for unpolarized initial and final
electrons (averaging over the initial electron spin) has the form:
\be\label{eq:10}
L_{\mu\nu}=q^2g_{\mu\nu}+2(k_{1\mu}k_{2\nu}+k_{1\nu}k_{2\mu}).
\ee

The hadronic tensor $W_{\mu\nu}$, for unpolarized initial and final
protons can be written in the standard form, through two unpolarized
structure functions:
\be\label{eq:11}
W_{\mu\nu}=\left ( -g_{\mu\nu}+\frac{q_{\mu}q_{\nu}}{q^2}\right )
W_1(q^2)+P_{\mu}P_{\nu} W_2(q^2).
\ee
Averaging over the initial proton spin, the structure functions
$W_i$, $i=1,2$, are expressed in terms of the nucleon
electromagnetic form factors:
\be\label{eq:12}
W_1(q^2)=-q^2G_M^2(q^2), \ \ W_2(q^2)=4M^2\frac{G_E^2(q^2)+\tau
G_M^2(q^2)}{1+\tau},
\ee
where $G_E(q^2)=F_1(q^2)- \tau F_2(q^2)$ is the proton electric form
factor and $\tau=-q^2/4M^2$.

The expression of the differential cross section, as a function of
the recoil-electron energy $\epsilon_2$, for unpolarized
proton-electron scattering can be written as:
\be\label{eq:13}
\frac{d\sigma^{(B)}}{d\epsilon_2}=\frac{\pi\alpha^2}{m\vec
p^2}\frac{\cal D}{q^4},
\ee
with
\be\label{eq:14}
{\cal D}=q^2(q^2+2m^2)G_M^2(q^2)+ 2\left
[q^2M^2+\frac{1}{1+\tau}\left(2mE+ \frac{q^2}{2}\right )^2\right
]\left [ G_E^2(q^2)+\tau G_M^2(q^2)\right ].
\ee
This expression is valid in the one-photon exchange (Born)
approximation in the reference system where the target electron is
at rest.

The expression of the differential cross section, as a function of
the four-momentum transfer squared, is
\be\label{eq:15}
\frac{d\sigma^{(B)}}{dq^2}=\frac{\pi\alpha^2}{2m^2\vec
p^2}\frac{\cal D}{q^4}.
\ee

And at last, the differential cross section over the
scattered-electron solid angle has the following expression
\be\label{eq:16}
\frac{d\sigma^{(B)}}{d\Omega_e}=\frac{\alpha^2}{8m^4|\vec{p}|}
\left (1-\frac{4m^2}{q^2}\right )^{3/2}\frac{\cal D}{E+m}.
\ee

\section{Radiative corrections}
Let us consider the model-independent QED radiative corrections
which are due to the vacuum polarization and emission of the virtual and real (soft and hard) photons in
the electron vertex. The corresponding diagrams are shown in Fig.\ref{diagrams}.
\subsection{Soft photon emission}
In this section we calculate the contribution to the radiative
corrections of the soft photon emission  when the photons are
emitted by the initial and final electrons
\be\label{eq:17}
p(p_1)+e(k_1)\to p(p_2)+e(k_2)+\gamma (k).
\ee
The matrix element in this case (the photon emitted from the
electron vertex) is given by
\be\label{eq:18}
{\cal
M}^{(\gamma)}=\frac{1}{q^2}(4\pi\alpha)^{3/2}j_{\mu}^{(\gamma)}J_{\mu},
\ee
where the electron current corresponding to the photon emission is
\be\label{eq:19}
j_{\mu}^{(\gamma)}=\bar{u}(k_2)\left [\frac{1}{d_1}\gamma_{\mu}(\hat{k}_1-\hat{k}+m)
\gamma_{\rho}+\frac{1}{d_2}\gamma_{\rho}(\hat{k}_2+\hat{k}+m)
\gamma_{\mu}\right ]u(k_1)A^*_{\rho},
\ee
where $A_{\rho}$ is the polarization vector of the emitted photon
and $d_1=-2k\cdot k_1, d_2=2k\cdot k_2$.

The differential cross section of reaction (\ref{eq:17}) can be written
as
\be\label{eq:20}
d\sigma^{(\gamma)}=\frac{(2\pi)^{-5}}{32m|\vec{p}|}|{\cal
M}^{(\gamma)}|^2 \frac{d^3\vec k_2}{\epsilon_2}\frac{d^3\vec
p_2}{E_2}\frac{d^3\vec k}{\omega} \delta^4(k_1+p_1-k_2-p_2-k).
\ee
It is necessary to integrate over the photon phase space. Since the
photons are assumed to be soft,  then the integration over the
photon energy is restricted to $\omega\leq \bar\omega$. The
quantity $\bar\omega$ is determined by particular experimental
conditions and it is assumed that $\bar\omega$ is sufficiently small
to neglect the momentum $k$ in the $\delta $ function and in the
numerators of the matrix element ${\cal M}^{(\gamma)}$. In order to
avoid the infrared divergence, which occurs in the soft photon cross
section, a small fictitious photon mass $\lambda $ is introduced.

In the soft photon approximation, the matrix element (\ref{eq:18}) is
\be\label{eq:21}
{\cal M}^{(\rm soft)}=\sqrt{4\pi\alpha}\left (\frac{k_2\cdot A^*}{k\cdot
k_2}-\frac{k_1\cdot A^*}{k\cdot k_1}\right ){\cal M}^{(B)}.
\ee
The differential cross section (\ref{eq:20}), integrated over the soft photon
phase space, can be written as
\be\label{eq:22}
d\sigma^{(\rm soft)}=\delta^{(s)}d\sigma^{(B)},
\ee
where the radiative correction due to the soft photon emission is
\be\label{eq:23}
\delta^{(s)}=-\frac{\alpha}{4\pi^2}\int_\lambda^{\bar\omega}\sqrt{\omega^2-\lambda^2}d\omega \int d\Omega_k\Bigl
[\frac{m^2}{(k\cdot k_1)^2}+\frac{m^2}{(k\cdot
k_2)^2}-2\frac{k_1\cdot k_2}{k\cdot k_1 k\cdot k_1}\Bigr ].
\ee
Assuming $\bar{\omega}\ll m$ and using the results of the paper \cite{'tHooft:1978xw}, we can do the
integration and the expression for $\delta^{(s)}$ has the form
\ba
\delta^{(s)}&=&\frac{\alpha}{\pi}\left \{1-2\ln{\frac{2\bar\omega}{\lambda}}+\frac{\epsilon_2}{k_2}\left
[\ln{\frac{\epsilon_2+k_2}{m}}
\left ( 1+2\ln{\frac{2\bar\omega}{\lambda}} +\ln{\frac{\epsilon_2+k_2}{m}}+2\ln{\frac{m}{2k_2}}\right )- \right .\right . \nn\\
&& \left . \left .  -\frac{\pi^2}{6}+Li_2\left (\frac{\epsilon_2-k_2}{\epsilon_2+k_2}\right )
\right ]\right  \}, \label{eq:24}
\ea
where $k_2\equiv |\vec{k}_2|$ ($\vec{k}_2$ is the momentum of the recoil electron) and $Li_2(x)$ is the Spence (dilogarithm) function defined as
$$ Li_2(x)=-\int_0^x\frac{\ln{(1-t)}}{t}dt. $$

\subsection{Virtual photon emission}
In this section, we calculate the contribution to the radiative
corrections of the virtual photon emission in the electron vertex
(the electron vertex correction) and the vacuum polarization term.

The matrix element corresponding to this process can be written as
\be\label{eq:25}
{\cal M}^{(\rm virt)}=\frac{1}{q^2}4\pi\alpha
J_{\mu}\bar{u}(k_2)\Lambda_{\mu}(k_1,k_2)u(k_1),
\ee
where we introduce
\be\label{eq:26}
\Lambda_{\mu}(k_1,k_2)=\frac{2i\alpha}{(2\pi)^3}\int
\frac{d^4k}{k^2-\lambda^2}\frac{\hat{O}_{\mu}}{(k^2-2k\cdot
k_1)(k^2-2k\cdot k_2)}
\ee
and the matrix $\hat{O}_{\mu}$ is
\be\label{eq:27}
\hat{O}_{\mu}=4k_1\cdot
k_2\gamma_{\mu}-2(\hat{k}_1\hat{k}\gamma_{\mu}+\gamma_{\mu}\hat{k}\hat{k}_2)-
2\hat{k}\gamma_{\mu}\hat{k}.
\ee
The integration over the virtual-photon four-momentum $k$ leads to the
following expression for the function $\Lambda_{\mu}(k_1,k_2)$
\ba
\Lambda_{\mu}(k_1,k_2)&=&\frac{\alpha}{4\pi}\left \{\left
[\ln\frac{\Lambda^2}{m^2}+\frac{1}{2}+\int_0^1\frac{dx}{P^2_x}\left (4m^2-
\frac{3}{2}q^2+(q^2-2m^2)\left(\ln\frac{P^2_x}{m^2}+\ln\frac{m^2}{\lambda^2}\right )\right )\right  ]\right .\gamma_{\mu}\nn\\
&&
\left . +m\int_0^1\frac{dx}{P^2_x}
\sigma_{\mu\nu}q_{\nu}\right \},
\label{eq:28}
\ea
where $P^2_x=m^2-x(1-x)q^2$ and
$\Lambda $ is the cut  parameter which define the region of infinite
momenta of the virtual photon. Thus we avoid the ultraviolet
divergence. The regularized vertex function can be obtained by the
subtraction of the contribution
$$\Lambda_{\mu}(k_1,k_1)=\frac{\alpha}{4\pi}\gamma_{\mu}\left [\ln\frac{\Lambda^2}{m^2}
+\frac{9}{2}-2\ln\frac{m^2}{\lambda^2}\right ] $$
from the expression (\ref{eq:28}).
As a result, we have
\be\label{eq:29}
\Lambda^R_{\mu}(k_1,k_2)=\Lambda_{\mu}(k_1,k_2)-\Lambda_{\mu}(k_1,k_1)=
\frac{\alpha}{4\pi}(A\gamma_{\mu}+B\sigma_{\mu\nu}q_{\nu}),
\ee
where
\be\label{eq:30}
A=-4+2\ln\frac{m^2}{\lambda^2}+\int_0^1\frac{dx}{P^2_x}\left \{4m^2-
\frac{3}{2}q^2+(q^2-2m^2)\left [\ln\frac{P^2_x}{m^2}+\ln\frac{m^2}{\lambda^2}\right ]\right \},
 \ \ B=m\int_0^1\frac{dx}{P^2_x}.
\ee
As we limit ourselves to the calculation of the radiative corrections at the order of $\alpha
 $ in comparison with the Born term, it is sufficient to calculate
 the interference of the Born matrix element with ${\cal M}^{(virt)} $
\be\label{eq:31}
|{\cal M}|^2=|{\cal M}^{(B)}|^2+2Re[{\cal M}^{(\rm virt)}{\cal
M}^{(B)*}]=(1+\delta_1+\delta_2)|{\cal M}^{(B)}|^2,
\ee
where the term $\delta_1$ is due to the modification of the
$\gamma_{\mu}$ term in the electron vertex, and the term $\delta_2$
is due to the presence of the $\sigma_{\mu\nu}q_{\nu}$ structure in the electron vertex.

The integration over the $x$ variable in the expression (\ref{eq:30}) gives the
following results for the radiative corrections due to the
emission of the virtual photon in the electron vertex
\ba
\delta_1&=&\frac{\alpha}{\pi}\left \{-2+2\ln\frac{m}{\lambda}\left
[1-\frac{\epsilon_2}{k_2}\ln\left (\frac{\epsilon_2+k_2}{m}\right )\right
]+\frac{m+3\epsilon_2}{2k_2}\ln \left(\frac{\epsilon_2+k_2}{m}\right )-
\frac{1}{2}\frac{\epsilon_2}{k_2}\ln\left (\frac{Q^2}{m^2}\right )
\ln\left(\frac{\epsilon_2+k_2}{m}\right )+\right .
\nn\\
&&\left .+\frac{\epsilon_2}{k_2}
\left [-\ln\left (\frac {m+\epsilon_2}{k_2}\right )\ln\left (\frac{\epsilon_2+k_2}{m}\right)+
Li_2\left(\frac{\epsilon_2+k_2+m}{2\left (m+\epsilon_2\right )}\right )-
Li_2\left(\frac{\epsilon_2-k_2+m}{2\left (m+\epsilon_2\right )}\right ) \right  ]\right \}, \nn\\
\delta_2&=&4\frac{\alpha}{\pi}\frac{mM^2q^2}{k_2{\cal D}}\ln\left(\frac{\epsilon_2+k_2}{m}\right )
(G_E^2-2\tau G_M^2).
\label{eq:32}
\ea
The radiative correction due to the vacuum polarization is (the electron loop has been taken into account):
\be\label{eq:33}
\delta^{(\rm vac)}=\frac{2\alpha}{3\pi}\left\{ -\frac{5}{3}+4\frac{m^2}{Q^2}+
(1-2\frac{m^2}{Q^2})\sqrt{1+4\frac{m^2}{Q^2}}\ln\frac{\sqrt{1+4\frac{m^2}{Q^2}}+1}
{\sqrt{1+4\frac{m^2}{Q^2}}-1}\right\}.
\ee
 For small and large values of
the $Q^2$ variable we have
\ba
&&\mbox{If \ \ } Q^2\ll m^2,   \ \ \
\delta^{(\rm vac)}=\frac{2\alpha}{15\pi}\frac{Q^2}{m^2}, \nn\\
&&\mbox{If \ \ }  \ \  Q^2\gg m^2,   \ \ \
\delta^{(\rm vac)}=\frac{2\alpha}{3\pi}\left [-\frac{5}{3}+\ln\frac{Q^2}{m^2}\right ].\nn
\ea
Taking into account the radiative corrections given by Eqs. (\ref{eq:24}, \ref{eq:32},
\ref{eq:33}), we obtain the following expression for the differential cross
section:
\be\label{eq:34}
d\sigma^{(RC)}=(1+\delta_1+\delta_2+\delta^{(s)}+\delta^{(\rm vac)})d\sigma^{(B)}=(1+\delta_0+\bar\delta +\delta^{(\rm vac)})d\sigma^{(B)},
\ee
where the radiative corrections $\delta_0$ and $\bar\delta $ are
given by
\ba
\delta_0&=&\frac{2\alpha}{\pi}\ln\frac{\bar\omega}{m}\left
[\frac{\epsilon_2}{k_2}\ln\left (\frac{\epsilon_2+k_2}{m}\right )-1\right ],\nn\\
\bar\delta &=&\frac{\alpha}{\pi} \left \{
-1-2\ln2+ \frac{\epsilon_2}{k_2}
\left [\ln \left( \frac{\epsilon_2+k_2}{m}\right ) \left (1+
\ln \left ( \frac{\epsilon_2+k_2}{m} \right )
+2\ln  \left(\frac{m}{k_2}\right )+
\frac{m+3\epsilon_2} {2\epsilon_2}-
\right . \right . \right .
\nn\\
&&\left . -\ln\left (\frac{\epsilon_2+m}{k_2}\right)-
\frac{1}{2}\ln\left (\frac{Q^2}{m^2}\right )\right )+4m\frac{M^2q^2}{\epsilon_2{\cal D}}
\ln\left(\frac{\epsilon_2+k_2}{m}\right )(G_E^2-2\tau G_M^2)- \nn\\
&&
\left . \left .-\frac{\pi^2}{6}+Li_2\left (\frac{\epsilon_2-k_2}{\epsilon_2+k_2}\right)+
Li_2\left(\frac{\epsilon_2+k_2+m}{2(\epsilon_2+m)}\right)-
Li_2\left(\frac{\epsilon_2-k_2+m}{2(\epsilon_2+m)}\right )\right ]\right \}.
\label{eq:35}
\ea

We separate the contribution $\delta_0$ since it can be summed up in
all orders of the perturbation theory using the exponential form of
the electron structure functions \cite{Kuraev:1985hb}. To do this it is
sufficient to keep only the exponential contributions in the
electron structure functions. The final result can be obtained by
the substitution of the term $(1+\delta_0 )$ by the following term
\be\label{eq:36}
\left(\displaystyle\frac{\bar\omega}{m} \right)^{\beta}\displaystyle\frac{\beta}{2}\int_0^1x^{\frac{\beta}{2}-1}(1-x)^{\frac{\beta}{2}}dx,
\ee
where
$$\beta =\displaystyle\frac{2\alpha}{\pi}
\left [\frac{\epsilon_2}{k_2}\ln \left (\displaystyle\frac{\epsilon_2+k_2}{m}\right)-1 \right ].
$$

\subsection{Hard photon emission}

In this section we calculate the radiative correction due to the hard photon emission
by the initial and recoil electrons only (the model-independent part).
The contribution due to radiation from the initial and scattered
protons (the model-dependent part) requires a special consideration and
we leave it for other investigations. We consider the experimental
setup when only the energies of the scattered proton and final electron
are measured.

The differential cross section of the reaction (\ref{eq:17}), averaged over
the initial particle spins, can be written as
\be\label{eq:37}
d\,\sigma^{(h)}=\frac{\alpha^3}{32\,\pi^2}\,\frac{1}{m\,p}\,
\frac{L_{\mu\nu}^{(\gamma)}\,W_{\mu\nu}}{q_1^4}\frac{d^3k_2}{\epsilon_2}
\frac{d^3p_2}{E_2}\frac{d^3k}{\omega} \delta(p_1+k_1-p_2-k_2-k),
\ee
where $ q_1=k_1-k_2-k$ and the leptonic tensor has the following
form
\ba\label{eq:38}
L_{\mu\nu}^{(\gamma)}&=&A_0\tilde{g}_{\mu\nu}+A_1\tilde{k}_{1\mu}\tilde{k}_{1\nu}+
A_2\tilde{k}_{2\mu}\tilde{k}_{2\nu}+A_{12}(\tilde{k}_{1\mu}\tilde{k}_{2\nu}+
\tilde{k}_{1\nu}\tilde{k}_{2\mu})\,, \\
A_0&=&4\left [\frac{d_1}{d_2}+\frac{d_2}{d_1}-2q_1^2\left
(\frac{m^2}{d_1^2}+\frac{m^2}{d_2^2}+2\frac{k_1\cdot
k_2}{d_1d_2}\right )\right ],\
A_1=16\frac{q_1^2}{d_1d_2}-32\frac{m^2}{d_2^2},  \nn\\
A_2&=&16\frac{q_1^2}{d_1d_2}-32\frac{m^2}{d_1^2}, \
A_{12}=-32\frac{m^2}{d_1d_2}\,.
\nn
\ea
The hadronic tensor is defined by Eqs.(11\,,12) with the substitution $q\to q_1.$

\begin{figure}[t]
\centering
\includegraphics[width=0.4\textwidth]{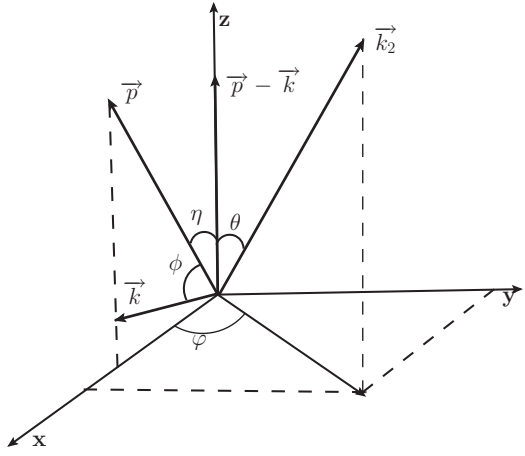}
 \parbox[t]{0.9\textwidth}{\caption{Coordinate system and definition of the angles
used for the integration over the variables of the final state.}
\label{angular space}}
\end{figure}

The contraction of leptonic and hadronic tensors reads
\be\label{eq:39}
L_{\mu\nu}^{(\gamma)}W_{\mu\nu}=-W_1(q_1^2)\,S_1 +\frac{W_2(q_1^2)}{M^2}\,S_2\,,
\ee
where the functions $S_{1,2}$ have the following expressions
\ba
S_1&=&8\left (\frac{d_1}{d_2}+\frac{d_2}{d_1}\right
)-\frac{16}{d_1d_2}(2m^2+q_1^2) \left [2k_1\cdot k_2+m^2\left
(\frac{d_1}{d_2}+\frac{d_2}{d_1}\right ) \right]\,,\label{eq:40}\\
S_2&=&4M^2\left [\frac{d_1}{d_2}+\frac{d_2}{d_1}-2q_1^2\left
(\frac{m^2}{d_1^2}+\frac{m^2}{d_2^2}+2\frac{k_1\cdot
k_2}{d_1d_2}\right )\right  ]+32\frac{m^2}{d_1d_2}(k\cdot p_1)^2+
16\frac{(k_1\cdot p_1)^2}{d_1}+
\nn\\
&&+16\frac{(k_2\cdot p_1)^2}{d_2}+16k_1\cdot p_1k_2\cdot p_1
\left [\frac{1}{d_1}+\frac{1}{d_2}-2\left
(\frac{m^2}{d_1^2}+\frac{m^2}{d_2^2}+2\frac{k_1\cdot
k_2}{d_1d_2}\right  )\right ]+\nn\\
&&+16k\cdot p_1\left  [\frac{k_1\cdot p_1}{d_2^2}(d_2-2m^2)-
\frac{k_2\cdot p_1}{d_1^2}(d_1-2m^2)+2\frac{k_1\cdot k_2}{d_1d_2}
(k_2\cdot p_1-k_1\cdot p_1)\right ]. \label{eq:41}
\ea
Integrating over the scattered proton variables we obtain the
following expression for the differential cross section
\be\label{eq:42}
d\sigma^{(h)}=\frac{\alpha^3}{32\pi^2}\frac{1}{m\,p} \int
\frac{d^3{k}}{\omega}\int \frac{d^3{k}_2}{\epsilon_2\,E_2}
\frac{1}{q_1^4}L_{\mu\nu}^{(\gamma)}W_{\mu\nu}\,\delta(m+E-\epsilon_2-E_2-\omega)\,.
\ee
To integrate further we have to define the coordinate system.
Following Ref. \cite{Kahane:1964zz}, where the $\pi-e^-$ scattering has
been analyzed, let us take the $z$-axis along the vector
$\vec{p}-\vec{k}$ and the momenta of the initial proton and emitted
photon lie in the $xz$ plane. The momentum of the scattered electron
is defined by the polar $\theta $ and azimuthal $\varphi $ angles as
it is shown in Fig.\ref{angular space}. The angle $\eta (\phi)$ is the angle
between the beam direction and $z$ axis (emitted photon momentum).

Integrating over the polar angle of the scattered electron we
obtain:
\be\label{eq:43}
\frac{d\sigma^{(h)}}{d\epsilon_2}=\frac{\alpha^3}{32\pi^2}\frac{1}{m\,p}
 \int \frac{d^3\,k}{\omega}\int \frac{d\varphi}{|\vec{p}-\vec{k}|}\,
\frac{1}{q_1^4}L_{\mu\nu}^{(\gamma)}\,W_{\mu\nu}.
\ee

\begin{figure}[t]
\centering
\includegraphics[width=0.4\textwidth]{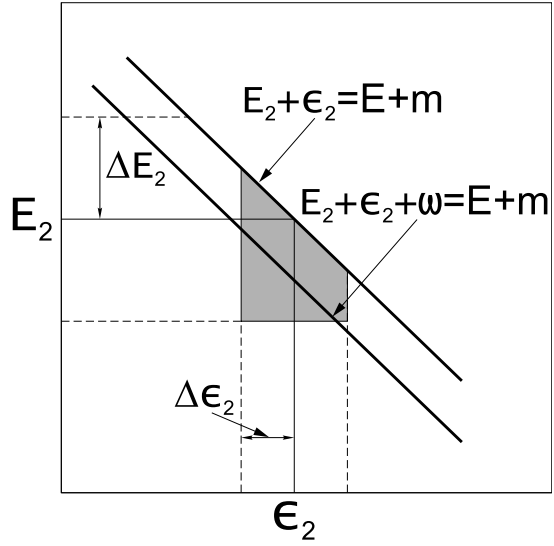}
 \parbox[t]{0.9\textwidth}{\caption{Plane of the $E_2$ and $\epsilon_2$
variables where the shaded area represents the kinematically allowed region within the experimental set-up.}\label{allowed events}}
\end{figure}

The region of allowed photon momenta should be determined. The experiment counts those events which, within the
accuracy of the detectors, are considered "elastic". We refer to the
experimental situation where the energies of the scatted proton and
recoil electron are measured. Because of the uncertainties in
determination of the recoil electron ($\Delta \epsilon_2$) and
scattered proton ($\Delta E_2$) energies, which usually are
proportional to $\epsilon_2$ and $E_2,$ respectively, the elastic
proton-electron scattering always accompanied by the hard photon
emission with the energies up to $\Delta \epsilon_2+\Delta E_2$.
At the proton beam energies of the order of 100 GeV this value can
reach a few GeV. The events for which the scattered proton energy is
$E_2\pm \Delta E_2$ and the recoil electron energy is $\epsilon_2
\pm \Delta \epsilon_2$ (they satisfy the condition
$E+m=E_2+\epsilon_2$) are considered as true elastic events. Here, $\Delta
E_2$ and $\Delta \epsilon_2$ are the errors in the measurement
of the final proton and recoil electron energies. The plot of the variable
$E_2$ versus the variable $\epsilon_2$ is shown in Fig.\ref{allowed events}. The
shaded area in this figure represents those events allowed by the
experimental limitations. The relation between the energies $E_2$
and $\epsilon_2$, as it is shown in Fig.\ref{allowed events}, has to be transformed
into a limit on the possible photon momentum $\vec{k}$. We
consider the experimental setup where no angles are measured and,
therefore, the orientation of the photon momentum
$\vec{k}$ is not limited.
In our calculation we restrict ourselves with the region $\epsilon_2<\epsilon_{2max}-\Delta E,$
where $\Delta E =\Delta E_2+\Delta\epsilon_2$ and $\epsilon_{2max}$ is defined by Eq.\,(5). In this case we get for the experimental
restriction the isotropic condition
$$ \omega \leq\Delta E.$$

In other case $\epsilon_2>\epsilon_{2max}-\Delta E$, the restriction for the photon energy is
$$ \omega \leq\epsilon_{2max}- \epsilon_2.$$

\begin{figure}[t]
\centering
\includegraphics[width=0.4\textwidth]{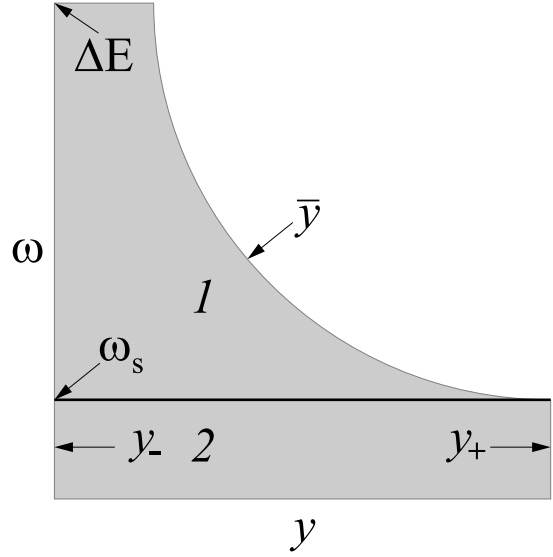}
 \parbox[t]{0.9\textwidth}{\caption{Integration region over the variables $\omega$
and $y$. Here $y_{\pm}=E\pm p, ~
\bar{y}=[(m-\epsilon_2)(E-\epsilon_2-\omega
)+\sqrt{\epsilon_2^2-m^2}\sqrt{(E+m-\epsilon_2-\omega )^2-M^2}]/
\omega .$ The quantity $\omega_s$ is defined by positive solution of the equation $\bar{y}=y_+$ and given by Eq. (\ref{eq:45}).}\label{integration region}}
\end{figure}

\begin{figure}[t]
\centering
\includegraphics[width=0.6\textwidth]{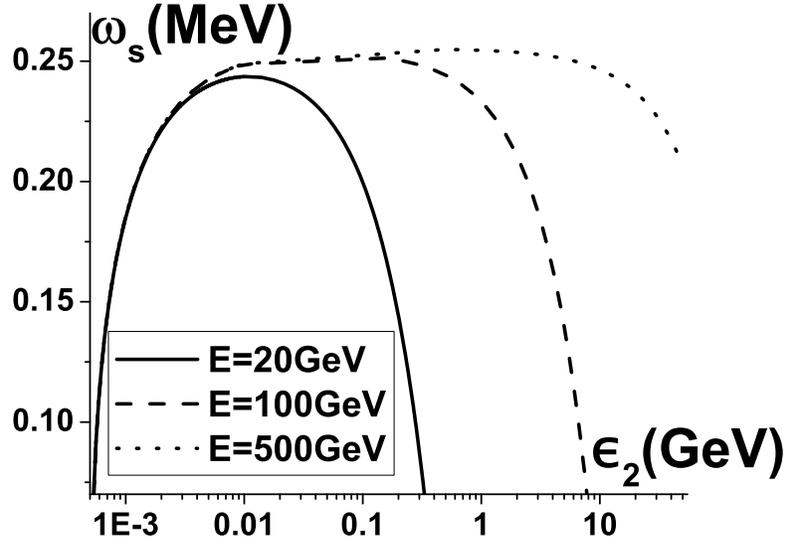}
 \parbox[t]{0.9\textwidth}{\caption{Maximum energy of the photon, when emitted in the whole angular phase space.}\label{omegas}}
\end{figure}

In the chosen coordinate system the element of solid angle becomes: $d^3 k\to
2\pi\,\omega^2\,d\omega\,d\cos{\phi}.$ Now we introduce a new
variable $y=E-p\cos{\phi}>0$ and rewrite Eq.~(\ref{eq:43}) as
\be\label{eq:44}
\frac{d\sigma^{(h)}}{d\epsilon_2}=\frac{\alpha^3}{16\,\pi}\frac{1}{m\,p^2}\int
\omega\,d\omega\int d\,y\int\limits_0^{2\pi}
\frac{1}{q_1^4\,|\vec{p}-\vec{k}|}
\Big(-W_1(q_1^2)\,S_1+\frac{W_2(q_1^2)}{M^2}\,S_2\Big)d\,\varphi\,,
\ee
where the integration region over the variables $\omega$ and $y$ is
shown in Fig.\ref{integration region}, and
\be\label{eq:45}
\omega_s=(|\vec{k}_2|-|\vec{p}|+x)\frac{M^2|\vec{k}_2|(|\vec{k}_2|+|\vec{p}|)
+(m-\epsilon_2)[M^2(x-m)+2m(2Ex+m^2-m\epsilon_2)]}{4(m-\epsilon_2)[x(M^2+m^2+2mE)-
m(M^2+mE)]+M^4},
\ee
where $x=E+m-\epsilon_2.$
The quantity $\omega_s$ represents the maximal energy, when the photon can be emitted in the whole angular
phase space. The dependence of this quantity on the recoil electron energy, at different values of the proton beam energy,
is shown in Fig.\ref{omegas}. We see that it is of the order of the electron mass $m$ in a wide range of the energies $\epsilon_2$ and $E$.  Because
our analytical calculations for the soft photon correction were performed under the condition $\bar\omega\ll m,$
where $\bar{\omega}$ is the maximal energy of the soft photon,
we can not
identify $\omega_s$ with $\bar\omega\ll m,$ as it has been done in the paper \cite{Kahane:1964zz}.

Following the Ref. \cite{Kahane:1964zz}, we include in the integral (\ref{eq:44}) the
weight function $g(\omega)$ given by
$$g(\omega)=1 \ \ \mbox{for} \ \ \bar{\omega}<\omega <\Delta E_2-\Delta\epsilon_2,$$
$$g(\omega)=\frac{\Delta E_2+\Delta\epsilon_2-\omega}{2\Delta\epsilon_2}
\ \ \mbox{for} \ \ \Delta E_2-\Delta\epsilon_2<\omega <\Delta
E_2+\Delta\epsilon_2\,.$$
In fact, the function $g(\omega)$ is the ratio of the straight line segments cut by the lines $E_2+\epsilon_2+\omega=E+m$
and $E_2+\epsilon_2=E+m$ in the shaded region in Fig.\ref{allowed events}.

So, the expression for the cross section given by Eq. (\ref{eq:44}) can be
written as a sum of two terms
\be\label{eq:46}
\frac{d\sigma^{(h)}}{d\epsilon_2}=\frac{\alpha^3}{16\,\pi}\frac{1}{m\,p^2}\bigg[\int
\limits_{\omega_s}^{\Delta E}g(\omega)\,C_1(\omega)\,d\omega +\int
\limits_{\bar{\omega}}^{\omega_s}C_2(\omega)\,d\omega\bigg]\,,
\ee
where
\ba\label{eq:47}
C_1(\omega)&=& \int\limits_{y_-}^{\bar{y}}\int\limits_0^{2\pi}\left [
\frac{\omega}{q_1^4\,|\vec{p}-\vec{k}|}
\left(-W_1(q_1^2)\,S_1+\frac{W_2(q_1^2)}{M^2}\,S_2\right)\right ]d\varphi
dy, \nn\\
C_2(\omega)&=&\int\limits_{y_-}^{y_+}\int\limits_0^{2\pi}\left(
\frac{\omega}{q_1^4\,|\vec{p}-\vec{k}|}
\left (-W_1(q_1^2)\,S_1+\frac{W_2(q_1^2)}{M^2}\,S_2\right )\right ]d\varphi
dy.
\ea

The scalar products of various 4-momenta, which enter in the
expressions for $S_1, S_2$ and $q_1^2$, are expressed, in terms of
the angles, as illustrated in Fig.\ref{angular space}, and the photon energy, as follows:
\ba\label{eq:48}
d_1&=&-2m\,\omega\,, \ \ k_1\cdot k_2=m\,\epsilon_2\,,
 \ k_1\cdot p_1=m\,E\,, \ \ k\cdot p_1=\omega(E-p\cos{\phi})\,,\nn\\
 d_2&=&2\,\omega[\epsilon_2-|\vec{k}_2 |(\cos{\theta}\,
\cos{(\eta+\phi)}+\cos{\varphi}\sin{\theta}\,\sin{(\eta+\phi)})]\,, \\
k_2\cdot p_1&=&\epsilon_2\,E-p|\vec{k}_2|(\cos{\eta}\cos{\theta}+
\cos{\varphi}\sin{\eta}\sin{\theta})\,.\nn
\ea
In turn, the respective trigonometric functions of angles are
expressed through the photon energy and the variable $y,$ as:
\ba\label{eq:49
}
\cos{\eta}&=&\frac{p^2-\omega(E-y)}{p|\vec{p}-\vec{k}|}\,,
\ \ \cos{(\eta+\phi)}=\frac{E-\omega-y}{|\vec{p}-\vec{k}|}\,,\nn\\
\cos{\theta}&=&\frac{(\epsilon_2-m)(E+m)+
\omega(y+m-\epsilon_2)}{|\vec{p}-\vec{k}|}\,, \ \ \sin{\theta}\,,
\ \sin{\eta}\,, \ \sin{(\eta+\phi)}\geq 0\,, \\
|\vec{p}-\vec{k}|&=&\sqrt{p^2+\omega(2y-2E+\omega)}\,.\nn
\ea
The functions $W_1$ and $W_2$ depend on the azimuthal angle $\varphi$, and, in order  to perform
the integration over this variable in the r.h.s. of Eq.(\ref{eq:46}),  one needs to use a specific expressions for the form factors
entering these functions. Further we concentrate on small values of the squared momentum transfer
as compared with the proton mass, where the form factors can be expanded in a series in term of powers of $q_1^2.$
In the calculations we keep the terms of the order of $1\,,$  $q_1^2$, and $q_1^4$ in the quantity
$$-W_1(q_1^2)\,S_1+\frac{W_2(q_1^2)}{M^2}\,S_2$$
which enters the differential cross section.

The integration in the r.h.s. of Eq. (\ref{eq:46}) over the $\varphi$ and $y$ variables is performed analytically.
The result for both $C_1(\omega)$ and $C_2(\omega)$ is very cumbersome, and it was published in the Appendix of our preprint \cite{Gakh:2016st}.
In the limit $\omega\to 0$ the function $C_1(\omega)$ is regular, and the function $C_2(\omega)$ has an infrared behaviour. We
extract the regular part $C_{2R}(\omega)$ and the infrared contribution $C_{2I}(\omega)$ by a simple subtraction procedure, by writing
\be\label{eq:50}
C_2(\omega)=\big[C_2(\omega)-C_2(\omega\to 0)\big]+C_2(\omega\to 0)=C_{2R}(\omega)+C_{2I}(\omega)\,, \ \ C_{2I}(\omega)\sim\frac{1}{\omega}
\left[\frac{\epsilon_2}{k_2}\ln\frac{\epsilon_2+k_2}{m}-1\right]\,.
\ee
The infrared contribution is combined with the correction due to soft and virtual photon emission and this results in change $\bar\omega\to\omega_s$ in
the expression for $\delta_0$ (see Eq. (\ref{eq:35})). The integration of the regular part $C_{2R}(\omega)$ over $\omega$ (as the lower limit we can chose an arbitrary small value) as well the whole contribution of the region 1, $C_1(\omega),$ is performed numerically.
\section{Numerical estimations and discussion}
In the  following section  the conditions for the experimental uncertainties are set to: $\Delta E_2=0.02E$ and $\Delta\epsilon_2=0.03\epsilon_2$
if other choice is not specified.

Since the four-momentum transfer squared is very small in this
reaction,  the proton charge and magnetic form factors
are approximated by Taylor series expansions. We use the expansion over the
variable $q^2$ of three form factor parameterizations.

\underline{By means of the radii (labeled as (r))}. In this approach we use the
expansion taking into account only the mean square radii that are
determined from the paper \cite{Lee:2015jqa}. These radii have been obtained
as a result of a comprehensive analysis of the electron-proton
scattering data (high statistics Mainz data set) using
model-independent constraints from the form factor analyticity. The
expansion is defined as follows
\be\label{eq:51}
\frac{G_{E,M}(q^2)}{G_{E,M}(0)}=1+\frac{1}{6}q^2r^2_{E,M}+O(q^4)\,,
\ee
where $r_{E,M}$ is proton electromagnetic charge (magnetic)
radius and their values are \cite{Lee:2015jqa}:
$r_E=0.904(15)$~fm=4.58~GeV$^{-1}$, $r_M=0.851(26)$~fm=4.32~GeV$^{-1}$.  Thus, electric
and magnetic form factors are ($q^2$ in GeV$^{-2}$):
$$G_E=1+3.496~q^2, ~\ G_M=2.793+8.65~q^2. $$

\underline{The dipole fits}. In this approach we use two different dipole fits. The well-known standard one, labeled as (sd), uses both, the small- and large-$Q^2$ data,
\be\label{eq:52}
G_{E}(q^2)=G, \ G_{M}(q^2)=\mu_p G, \ G=(1-1.41 q^2)^{-2}\,,
\ee
leads to the following expansions of the form factors, up to the terms $q^4$,
$$G_E=1+2.82 q^2+5.96 q^4, \ G_M=2.793+7.88 q^2+16.65 q^4. $$

Another dipole fit
\cite{Horbatsch:2016smx}, labeled as (d) uses only the lower-Q$^2$ data by MAMI Collaboration
\be\label{eq:53}
G_{E}(q^2)=(1-1.517\,q^2)^{-2}, \ G_{M}(q^2)=\mu_p (1-1.37 q^2)^{-2}\,,
\ee
and gives
$$G_E=1+3.034 q^2+6.91 q^4, \ G_M=2.793+7.65 q^2+15.72 q^4. $$

\underline{The sum of monopole terms, labeled as (m) }. In this approach we use the five-parameter
fit for both Dirac and Pauli form factors as a sum of three monopoles \cite{Blunden:2005ew}
\be\label{eq:54}
F_1(q^2)=\sum_1^3\frac{n_i}{d_i-q^2}, \
F_2(q^2)=\sum_1^3\frac{m_i}{g_i-q^2},
\ee
where $n_i, m_i, d_i$ and $g_i$ are free parameters, and the
parameters $n_3$ and $m_3$ are determined from the normalization
conditions
$$F_1(0)=\sum_i\frac{n_i}{d_i}, \ F_2(0)=\sum_i\frac{m_i}{g_i}. $$
The parameters $n_i, m_i, d_i$ and $g_i$ for the $F_1$ and $F_2$ proton form
factors are given in Table I. The normalization conditions are
$F_1(0)=1$ and $F_2(0)=\mu_p-1$, where $\mu_p$ =2.793 is the proton
total magnetic moment.

Thus, we have for the parameters $n_3$ and $m_3$
$$n_3=d_3-d_3\left (\frac{n_1}{d_1}+\frac{n_2}{d_2}\right ), \
m_3=g_3(\mu_p-1)-g_3\left (\frac{m_1}{g_1}+\frac{m_2}{g_2}\right ). $$

The expansions for the form factors $G_{E,M}$ are
\ba\label{eq:55}
G_E&=&1+\left [\sum\frac{n_i}{d_i^2}+\frac{\mu_p-1}{4m_p^2}\right ]q^2+
\left [\sum\frac{n_i}{d_i^3}+\frac{1}{4m_p^2}\sum\frac{m_i}{g_i^2}\right ]q^4\,,  \nn\\
G_M&=&\mu_p+\left [\sum\frac{n_i}{d_i^2}+\sum\frac{m_i}{g_i^2}\right ]q^2+
\left [\sum\frac{n_i}{d_i^3}+\sum\frac{m_i}{g_i^3}\right ]q^4\,.
\ea

The expansion of the form factors is as follows
$$G_E=1+3.017\,q^2+7.22\,q^4\,, \ \ G_M=2.793+8.239\,q^2+20.31\,q^4\,.$$

The d- and m-parameterizations give very close distributions, and therefore, we use only m-parametrization in our numerical calculations.

\begin{table}
\caption{\label{tab:table1}Parameters for the proton form factor fits in Eq. (\protect\ref{eq:54}) used
in this work, with $n_i, m_i, d_i$ and $g_i$ in units of GeV$^2$. }
\begin{tabular}{|c|c|c|c|}
  \hline
   $n_1$ & 0.38676 & $m_1$ & 1.01650 \\
   $n_2$ & 0.53222 & $m_2$ & -19.0246 \\
   $d_1$ & 3.29899 & $g_1$ & 0.40886 \\
   $d_2$ & 0.45614 & $g_2$ &  2.94311 \\
   $d_3$ & 3.32682 & $g_3$ & 3.12550 \\
  \hline
\end{tabular}
\end{table}

\underline{One-parameter linear model in conformal mapping variable labeled as (z)}. This approach is to use an expansion in $q^2$ of the approximation to the form factors given by one-parameter formulas
\be\label{eq:56}
G_E^2=1-C_E\,z\,, \ \ G_M^2=\mu_p^2(1-C_M\,z)\,, \ \ z=\frac{\sqrt{4m_{\pi}^2-q^2}-\sqrt{4m_{\pi}^2}}{\sqrt{4m_{\pi}^2-q^2}+\sqrt{4m_{\pi}^2}}\,,
\ee
with
\cite{Horbatsch:2016smx}
$$C_E=2.105\,, \ \ C_M=2.04\,.$$
In this case the expansion for the form factors reads
$$G_E=1+3.018\,q^2+7.221\,q^4\,, \ \ G_M=2.793+9.133\,q^2+43.65\,q^4\,.$$

To understand better the small-$Q^2$ distribution and its dependence on $r_E^2$, we expand function ${\cal D}$ defined by Eq. (\ref{eq:14}) for radii (r) (when the electric form factor
is smallest) and monopole (when the electric form factor is middle) parameterizations at two values of the proton beam energy: 100 and 500 GeV
\ba\label{eq:57}
{\cal D}(r,~E=100\,GeV )=0.0209+(1.92+0.00582\,r_E^2)q^2+(3.71+0.657\,r_E^2+0.00058\,r_E^4)q^4\,, \nn \\
{\cal D}(m,~E=100\,GeV )=0.0209+(1.92+0.00696\,r_E^2)q^2+(3.74+0.657\,r_E^2+0.00145\,r_E^4)q^4\,, \nn \\
{\cal D}(r,~E=500\,GeV )=0.5222+(1.772+0.174\,r_E^2)q^2+(-5.031+0.977r_E^2+0.014\,r_E^4)q^4\,, \nn \\
{\cal D}(m,~E=500\,GeV )=0.5222+(1.772+0.174\,r_E^2)q^2+(-4.203+0.977r_E^2+0.036\,r_E^4)q^4\,,
\ea
where $r_E^2$ must be taken in GeV$^{-2}.$

There is a compensation of the first two terms of these expansions when $|q^2|$ increases, and since the coefficient in front of $q^4$ is large, it have to be taken into account even at small enough values of $|q^2|$. The coefficient in front of $r_E^2$ in the second term increases rapidly with the growth of the beam energy.

\begin{figure}[t]
\centering
\includegraphics[width=0.5\textwidth]{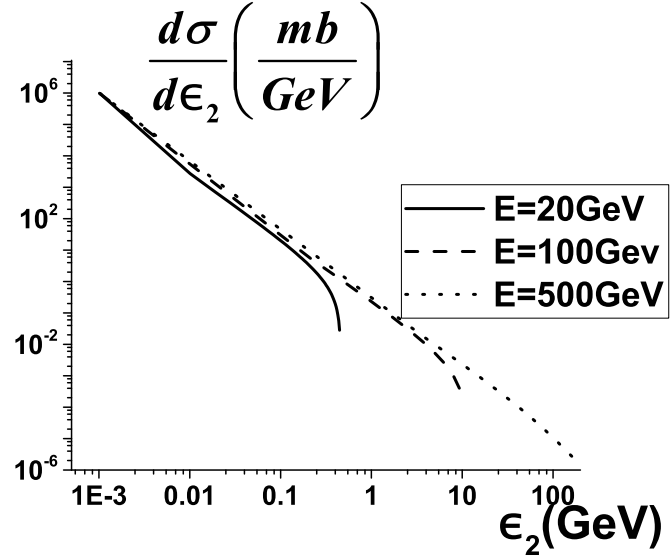}
 \parbox[t]{0.9\textwidth}{\caption{Born differential cross section, defined by Eq. (\ref{eq:13}), is calculated with the standard dipole fit of the form factors at different beam energies.}\label{elastic}}
\end{figure}

To illustrate the dependence of the recoil electron distribution on the  proton beam energy, the Born cross section is shown in Fig.\ref{elastic}, for the standard dipole fit
at $E=$20 GeV, 100 GeV  and  500 GeV. Here and further for the beam energy 500 GeV we restrict the recoil electron energy by 50 GeV, because for larger values the above expansions of the form factors are incorrect.

\begin{figure}[h]
\centering
\includegraphics[width=0.3\textwidth]{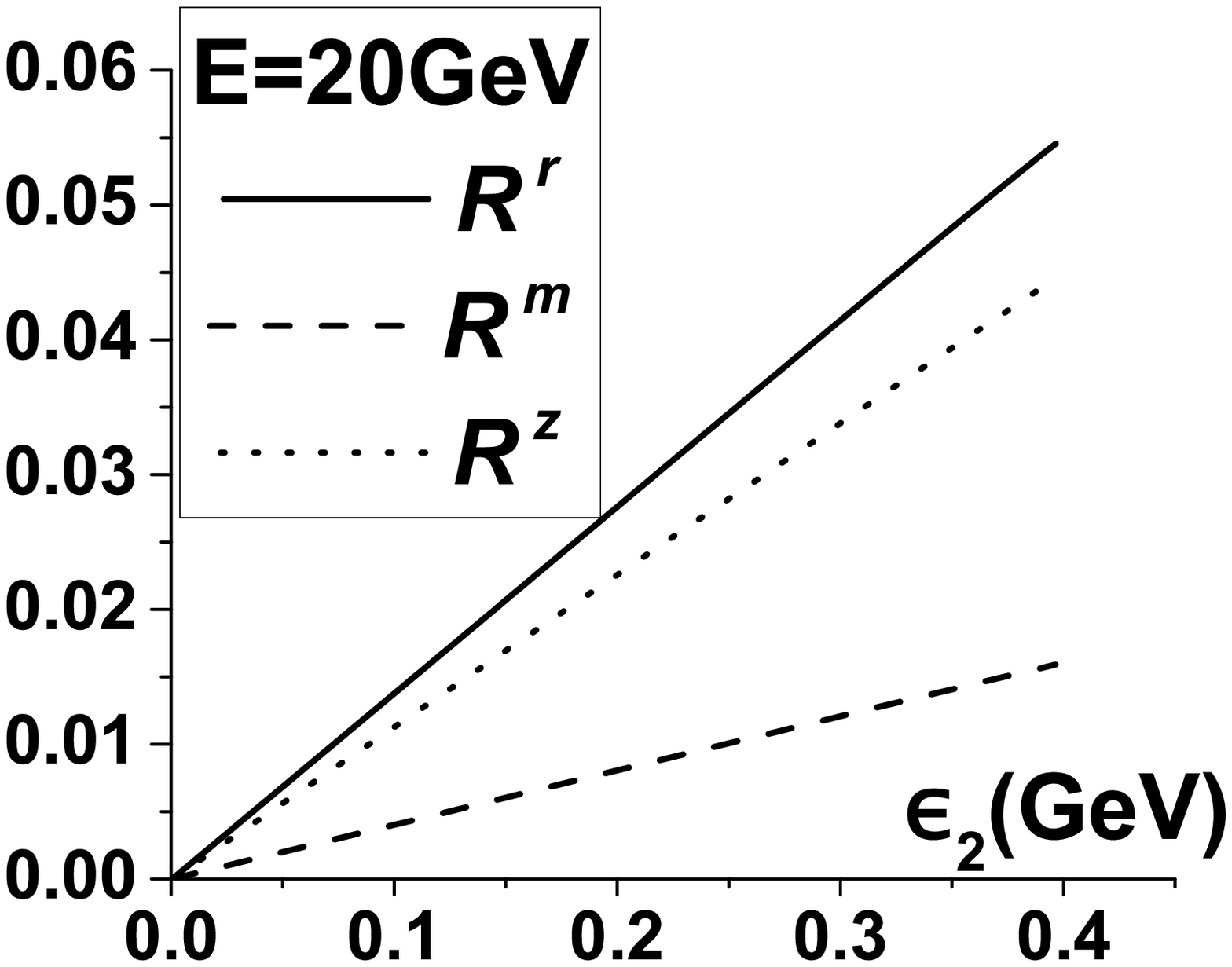}
\includegraphics[width=0.3\textwidth]{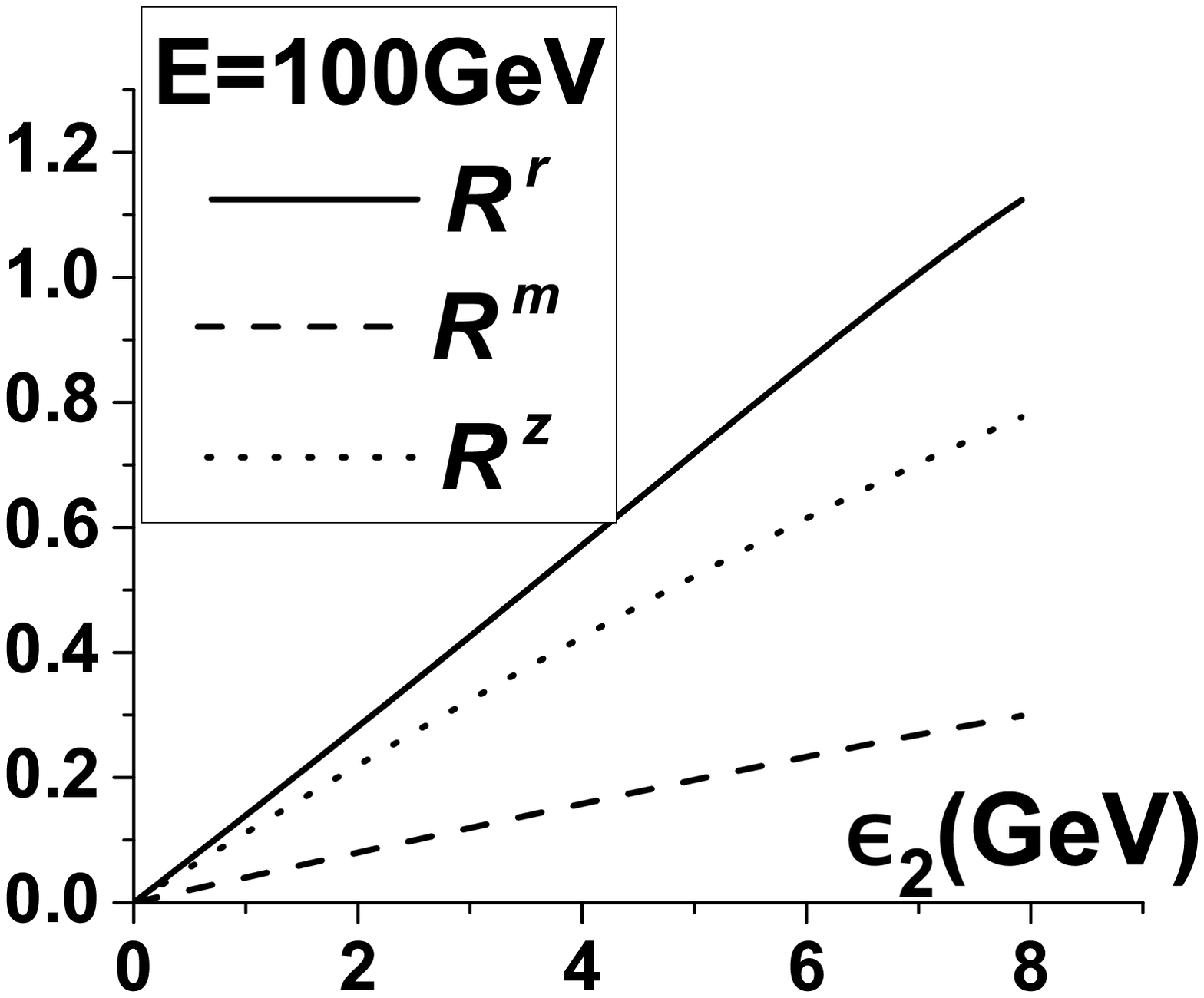}
\includegraphics[width=0.3\textwidth]{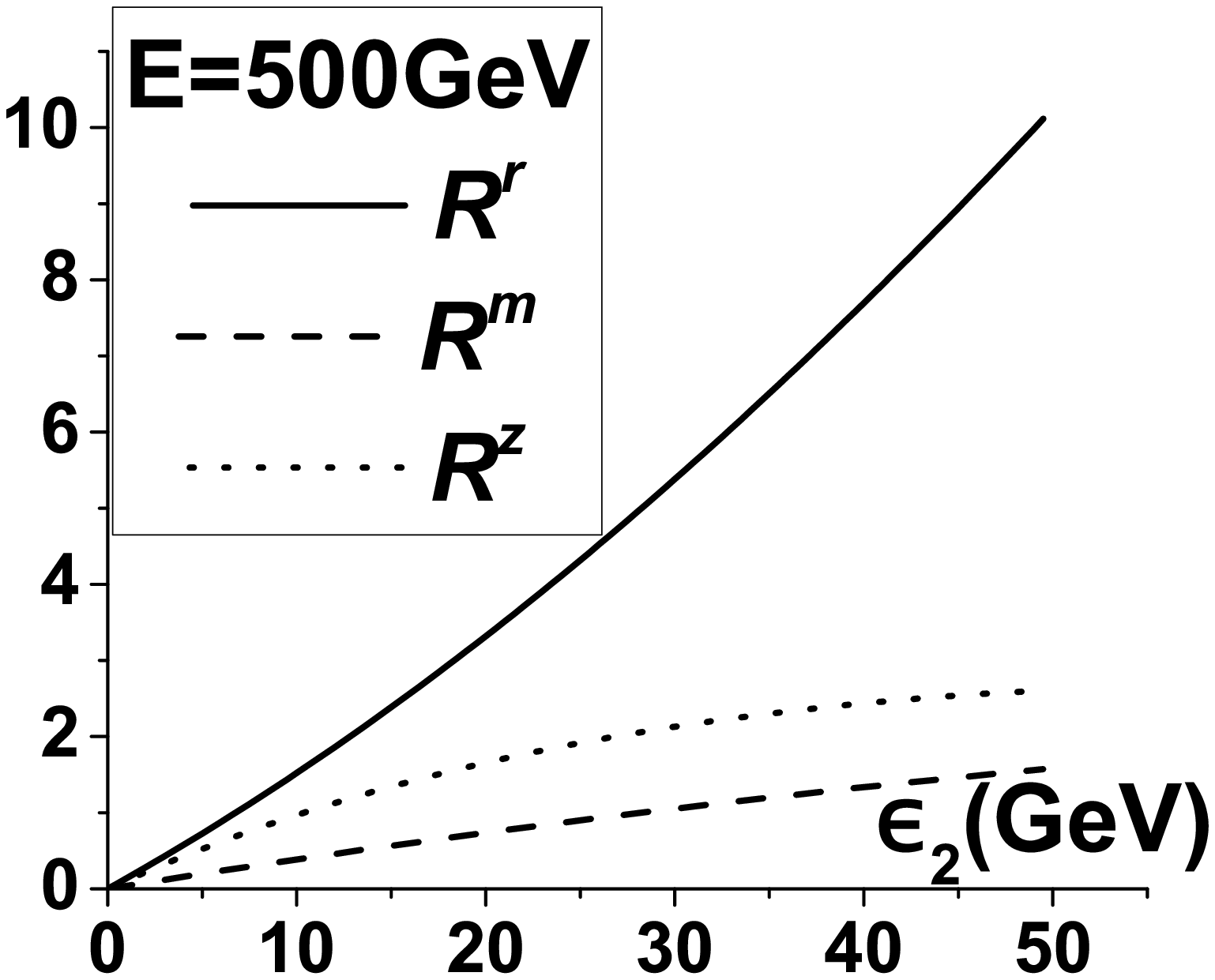}
\parbox[t]{0.9\textwidth}{\caption{The difference of the recoil electron distributions (Eq.~(\ref{eq:58})) in percent for various  parameterizations of the form factors, at proton energies 20 GeV, 100 GeV and 500 GeV  .}\label{parameterization}}
\end{figure}

The sensitivity of this cross section to different form factor parameterizations is shown in Fig.\ref{parameterization}, in terms of the quantities
(in percent)
\be\label{eq:58}
R^r=1-\frac{d\,\sigma^r}{d\,\sigma^{sd}}\,, \ \ R^m=1-\frac{d\,\sigma^m}{d\,\sigma^{sd}}\,, \ \ R^z=1-\frac{d\,\sigma^z}{d\,\sigma^{sd}}\,,
\ee
where $d\,\sigma^i$ is the differential cross section (\ref{eq:13}), where the indices $i=r\,,~z\,,~m\,, d$ correspond to above mentioned parameterizations
of the proton form factors.

\begin{figure}[t]
\centering
\includegraphics[width=0.43\textwidth]{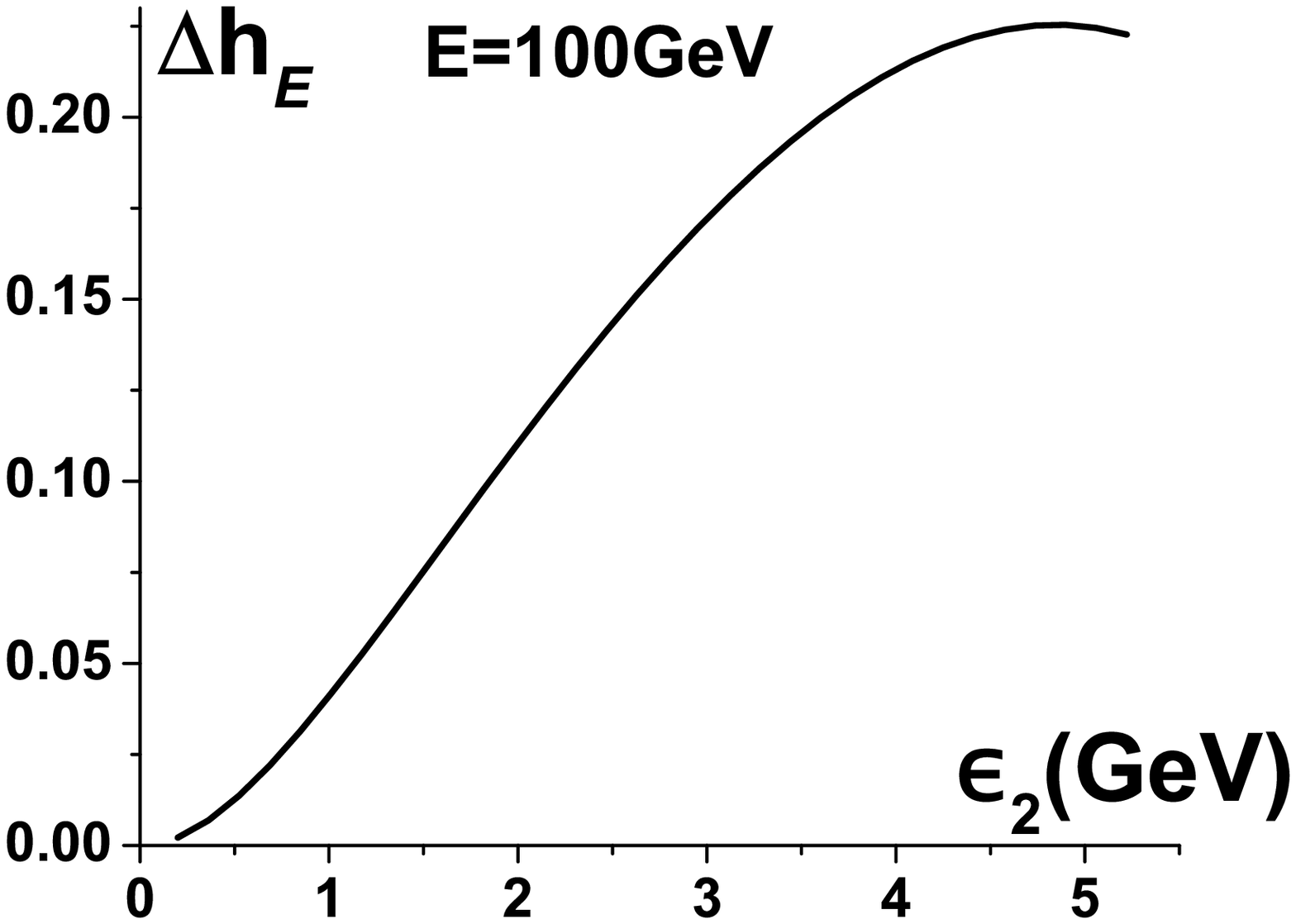}
\includegraphics[width=0.43\textwidth]{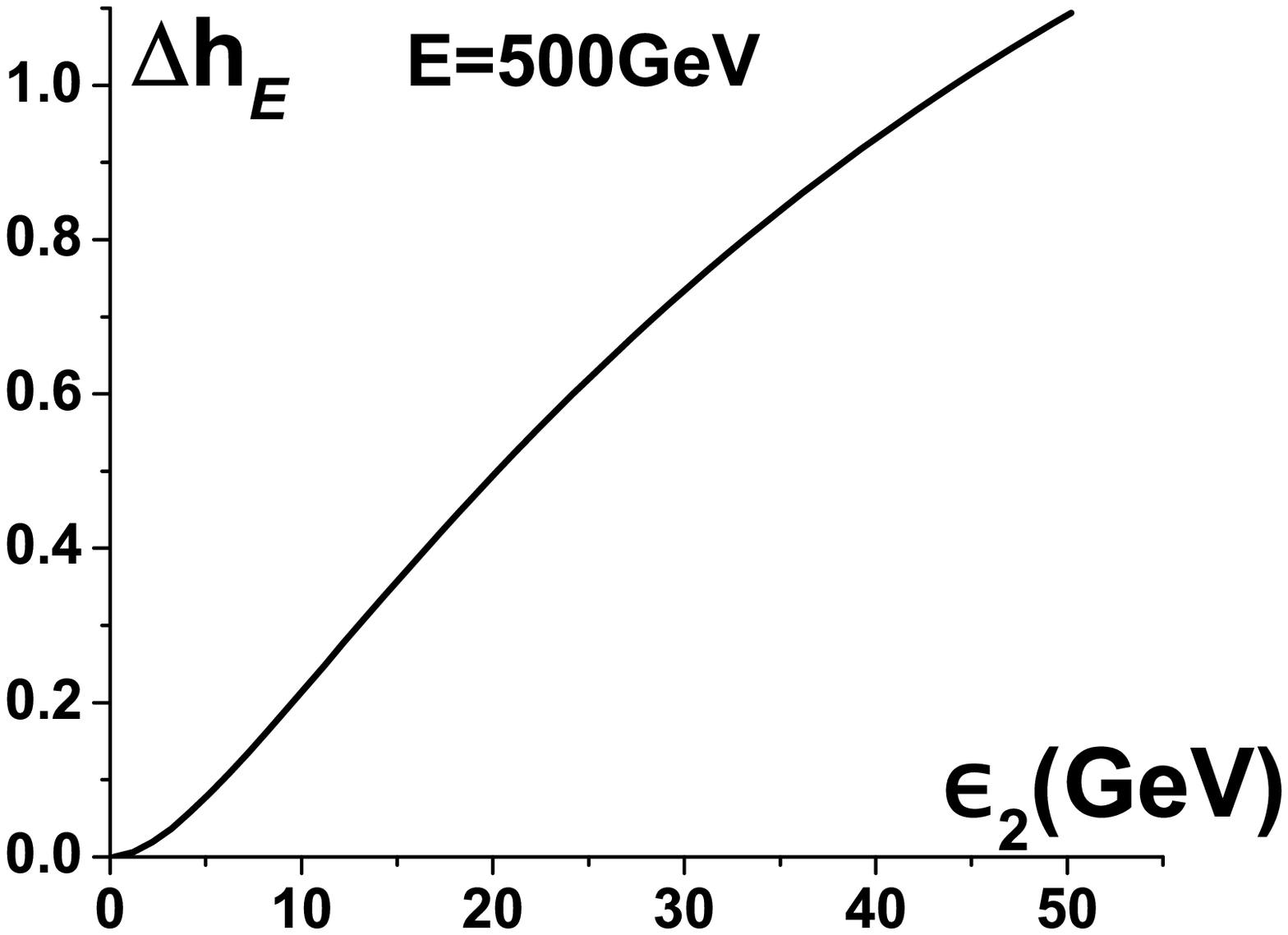}
\vspace{1cm}
\includegraphics[width=0.43\textwidth]{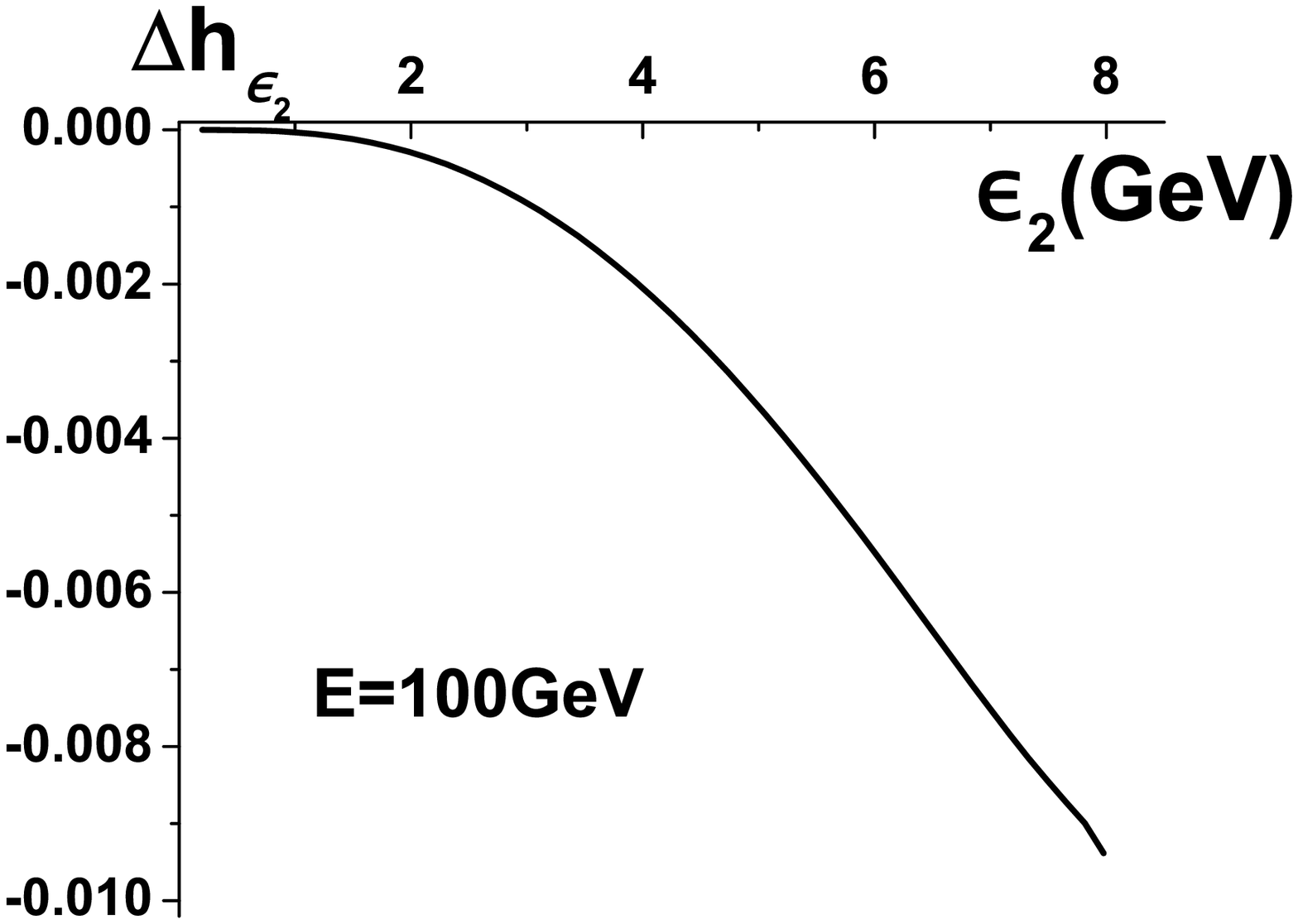}
\includegraphics[width=0.43\textwidth]{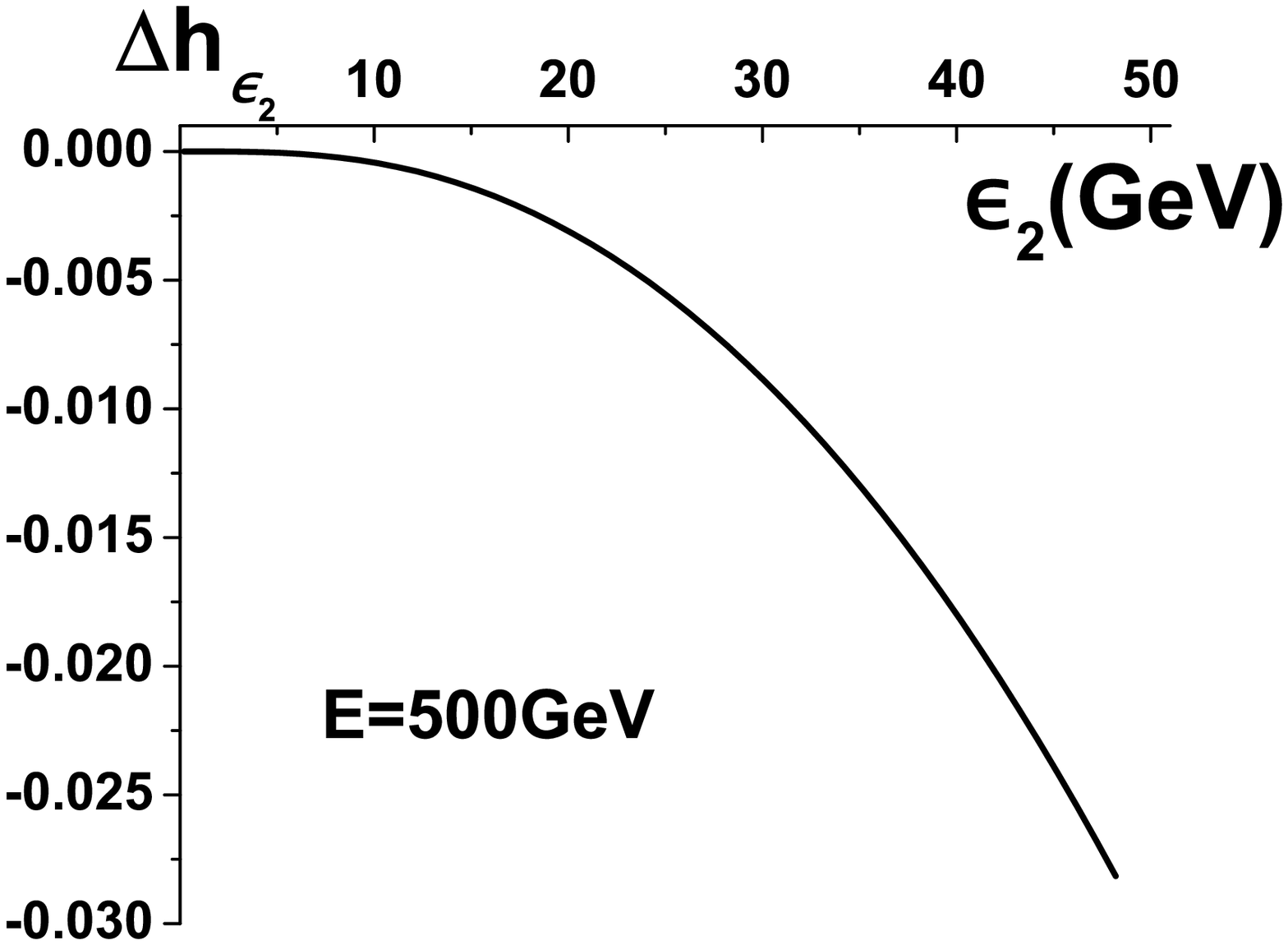}
\parbox[t]{0.9\textwidth}{\caption{The sensitivity, as the function of the recoil electron energy, (in percent) of the hard photon correction to the parameter $\Delta E_2$ (above) and $\Delta \epsilon_2$ (bottom) Eq.~(\ref{eq:59}) at  proton energy of 100 GeV  (left) and 500 GeV(right).}\label{Delta E}}
\end{figure}

The hard photon correction depends on the parameters $\Delta E_2$  and $\Delta \epsilon_2,$ due to the contribution of the region 1 in Fig.\ref{integration region}. To illustrate  this dependence, we show in Fig.\ref{Delta E} the quantities (in which the contribution of the region 2 is removed)
\ba\label{eq:59}
\Delta h_E=\frac{d\,\sigma^{(h)}(\Delta E_2=0.05\,E\,,~\Delta \epsilon_2=0.03\,\epsilon_2)}{d\,\sigma^{(B)}}-\frac{d\,\sigma^{(h)}(\Delta E_2=0.02\,E\,,~\Delta \epsilon_2=0.03\,\epsilon_2)}{d\,\sigma^{(B)}}\,, \nn \\
\Delta h_\epsilon=\frac{d\,\sigma^{(h)}(\Delta E_2=0.02\,E\,,~\Delta \epsilon_2=0.06\,\epsilon_2)}{d\,\sigma^{(B)}}-\frac{d\,\sigma^{(h)}(\Delta E_2=0.02\,E\,,~\Delta \epsilon_2=0.03\,\epsilon_2)}{d\,\sigma^{(B)}}\,,
\ea
as a function of the recoil electron energy for sd-parametrization. The cross section increases with the growth of $\Delta\,E_2$ at fixed value
$\Delta\,\epsilon_2,$ but even decreases with the growth of $\Delta\,\epsilon_2$ at fixed $\Delta\,E_2$. Such unusual dependence on the energy-cut parameters
is due to the weight function $g(\omega)$ in the integrand over the region $1$ in Fig.\ref{integration region}.
If $\Delta\,E_2$ increases then the region, where $g(\omega)=1,$ is enlarged. Meanwhile, the region, where $g(\omega)<1,$ is only shifted but the function $g(\omega)$ grows,
and these effects lead to the enhancement of the cross section. At increasing of $\Delta\,\epsilon_2$ the region, where $g(\omega)=1,$ is reduced, the region, where $g(\omega)<1,$ is enlarged and $g(\omega)$ decreases. The change of the cross section in the last case depends on interplay of these factors as well on the integrand.

Qualitatively, it can be understand if we change $C_1(\omega)$ in the r.h.s. of Eq.~(\ref{eq:46}) by its small-$\omega$ behaviour $\sim 1/\omega$ and performing analytical integration
\be\label{eq:60}
\int\limits_{\omega_s}^{\Delta E}g(\omega)\,\frac{d\omega}{\omega}=\ln\frac{\Delta\,E_2}{\omega_s}-\frac{1}{6}\left(\frac{\Delta\,\epsilon_2}{\Delta\,E_2}
\right)^2\,, \ \ \frac{\Delta\,\epsilon_2}{\Delta\,E_2}\ll\,1\,.
\ee
Thus, when parameters $\Delta\,E_2$ and $\Delta\,\epsilon_2$ grow, the logarithmic increase with $\Delta\,E_2$ and very weak decrease with
$\Delta\,\epsilon_2$ take place.

Note that our choice of parameters $\Delta E_2$  and $\Delta \epsilon_2$ is taken only for illustrative goal. Really, they have to be specific for every experiment, but our approach allows to calculate with any ones.

\begin{figure}[t]
\centering
\includegraphics[width=0.4\textwidth]{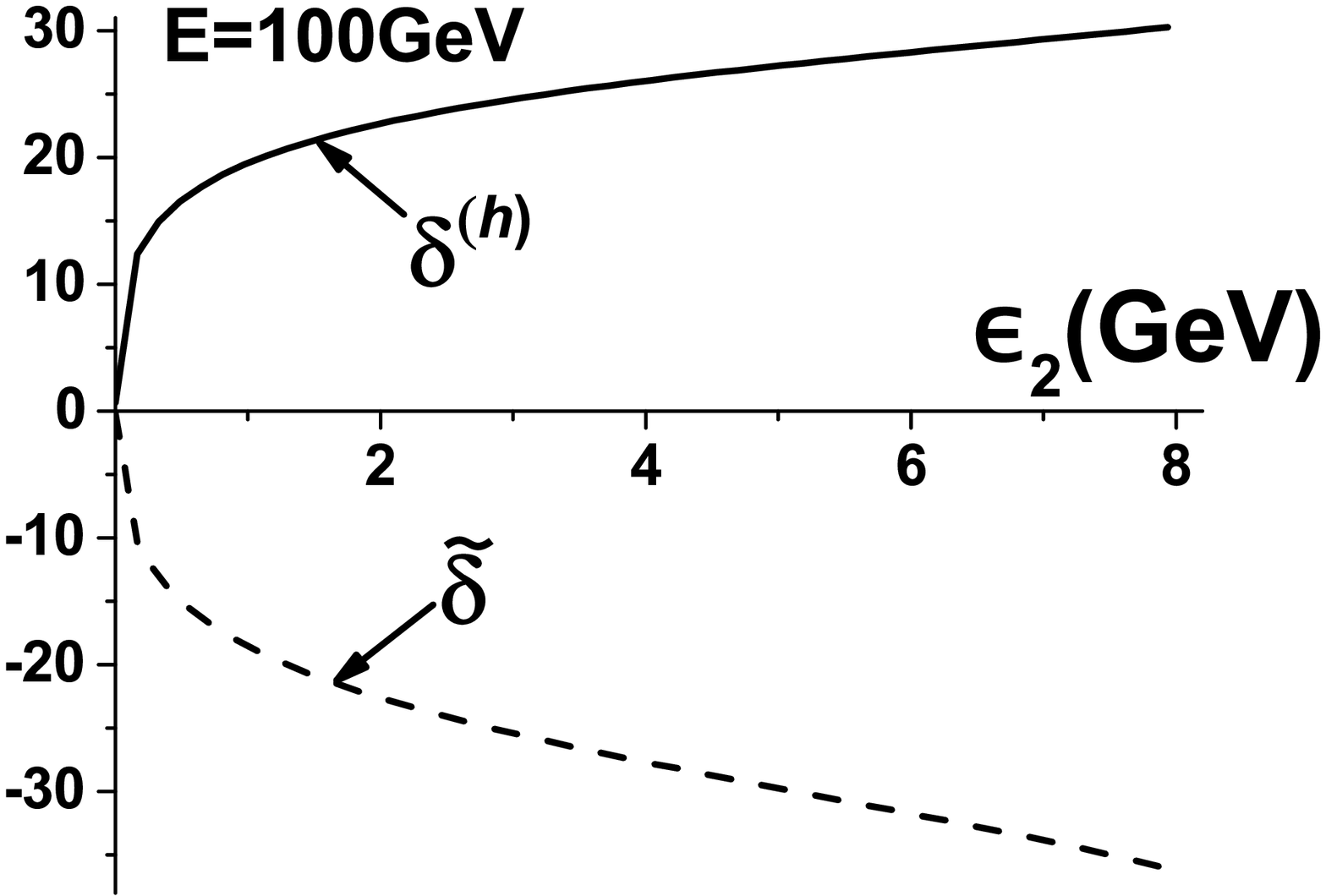}
\includegraphics[width=0.4\textwidth]{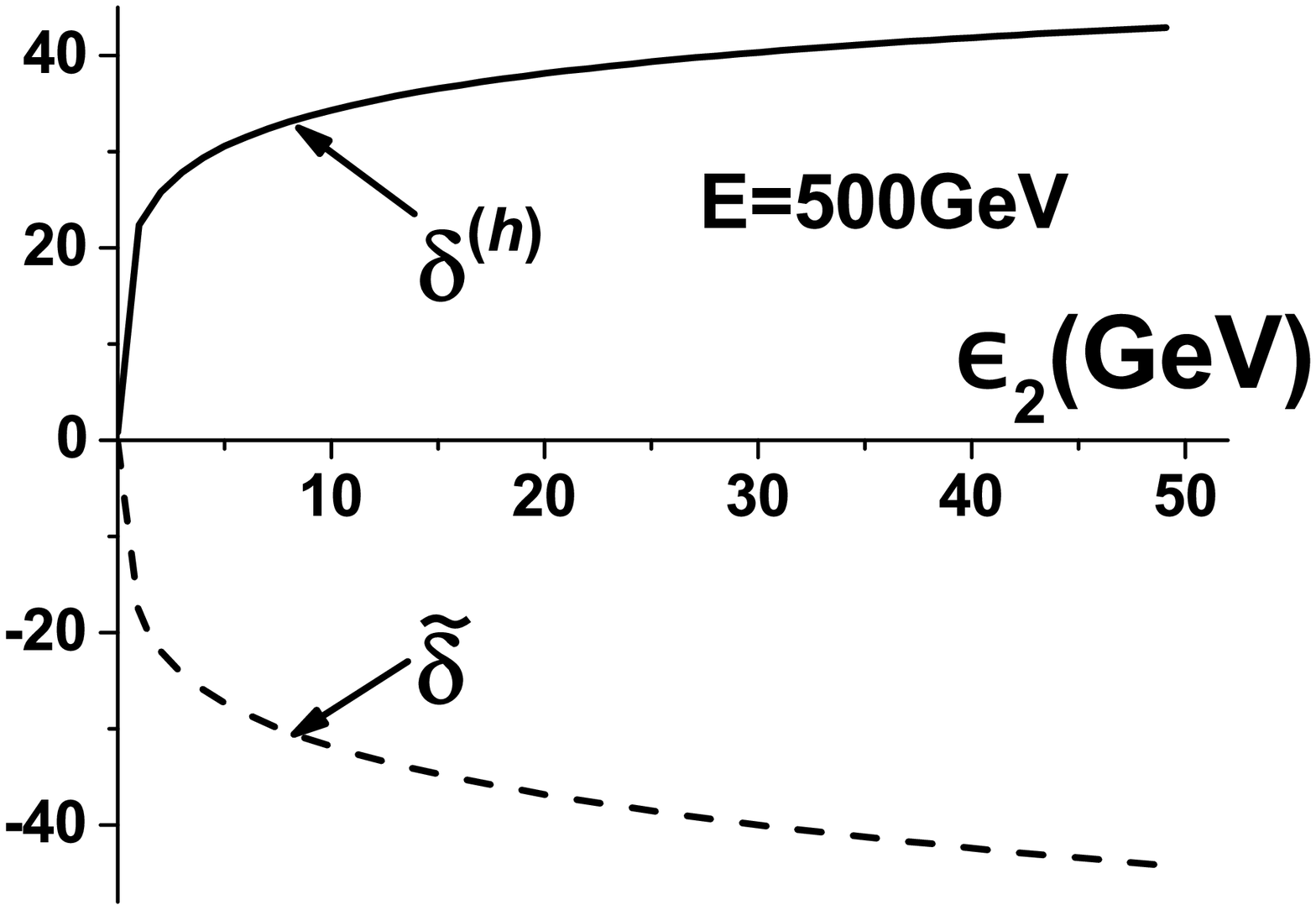}

\vspace{0.6cm}
\includegraphics[width=0.4\textwidth]{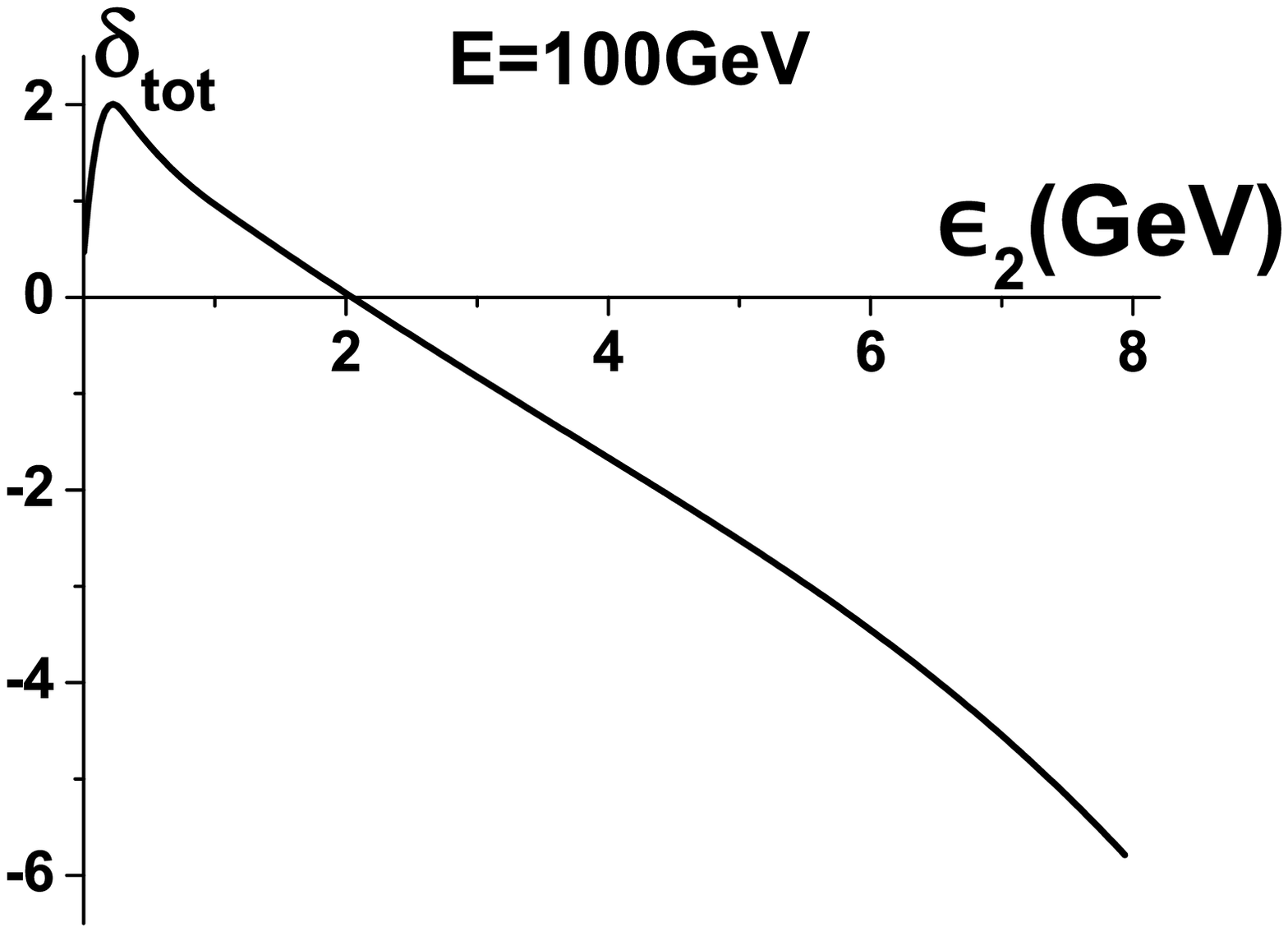}
\includegraphics[width=0.4\textwidth]{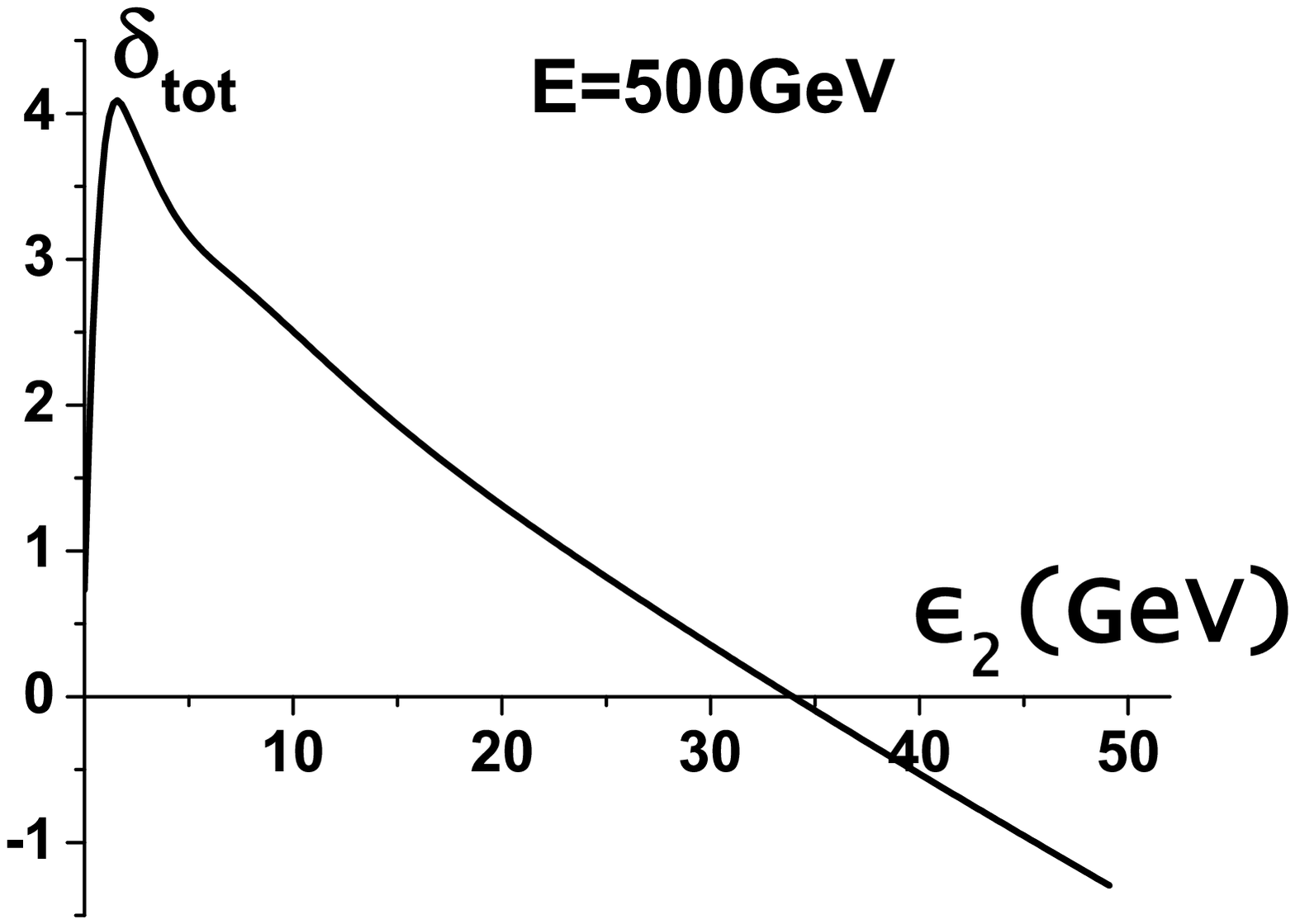}
 \parbox[t]{0.9\textwidth}{\caption{(Top) The modified soft and virtual ($\widetilde\delta$)  (dashed line) and hard $(\delta^{(h)})$ (solid line) corrections (in percent) as defined by Eq.(\ref{eq:61}). (Bottom)
 The total radiative correction (in percent) calculated for the standard dipole fit at $\Delta E_2 =0.02\,E$, at 100 GeV (left) and 500 GeV (right) incident proton energy.}\label{radiative corrections}}
\end{figure}

In Fig.\ref{radiative corrections} we present the quantities $\delta^{(h)}$ and $\widetilde{\delta},$
defined as
\ba\label{eq:61}
\delta^{(h)}=\frac{d\,\sigma^{(h)}}{d\,\sigma^{(B)}}-\frac{2\,\alpha}{\pi}\ln\frac{\omega_s}{\bar{\omega}}
\left[\frac{\epsilon_2}{k_2}\ln\left(\frac{\epsilon_2+k_2}{m}\right)-1\right ]\,, \nn \\ \widetilde{\delta}=\bar{\delta}+\delta^{(\rm vac)}+\frac{2\,\alpha}{\pi}\ln\frac{\omega_s}{m}
\left[\frac{\epsilon_2}{k_2}\ln\left(\frac{\epsilon_2+k_2}{m}\right )-1\right ]\,,
\ea
which we call  "modified hard and soft and virtual corrections", respectively,
as well their sum $\delta_{\rm tot}=\delta^{(h)}+\widetilde{\delta}$ that is the total model-independent first order radiative correction
(the last term in $\tilde{\delta}$ is $\delta_0(\bar{\omega}\to \omega_s)).$
In fact
$$\delta_{\rm tot}=\delta^{(h)}+\tilde\delta=\delta_{0}+\bar{\delta}+\delta^{(\rm vac)}+\frac{d\sigma^{(h)}}{d\sigma^{(B)}}\,.$$
Note, that both modified corrections in Eq.(\ref{eq:61}) are independent on the auxiliary parameter $\bar\omega$ but depend on the physical parameter $\omega_s$ and, therefore, have a physical sense.

To calculate
$\sigma_{\rm tot},$ we can write the quantity $(1+\delta_0(\bar{\omega}\to \omega_s))$
using the expression (\ref{eq:35}) or its exponential form defined by (\ref{eq:36}) (with substitution $\bar{\omega}\to \omega_s$). But numerical estimations show that they differ very insignificantly, by a few tenth of the percent, and further we do not use the exponential form.

We see that at small values of the squared momentum transfer (small recoil-electron energy $\epsilon_2$) the total model-independent radiative correction is positive and it decreases (with increase of $\epsilon_2$), reaching zero and becoming negative. The absolute value of the radiative correction does not exceed 6$\%,$ although the strong compensation of the large (up to 40 \%) positive "modified hard" and negative "modified soft and virtual" corrections takes place. Such behavior of the pure QED correction is similar to one derived in Ref.\,\cite{Kahane:1964zz}.

If the proton form factors are determined independently with high accuracy from other experiments, the measurement of the cross section $d\,\sigma/d\,\epsilon_2$ can be used, in principle, to
measure the model-dependent part of the radiative correction in the considered conditions. This possibility is similar to the one described in Ref.\,\cite{Abbiendi:2016xup} where the authors proposed to determine the hadronic (model-dependent) contribution to the running electromagnetic coupling $\alpha(q^2)$ by a precise measurement of the $\mu^--\,e^-$ differential cross section, assuming that QED model-independent radiative corrections are under control.

\begin{figure}[t]
\centering
\includegraphics[width=0.4\textwidth]{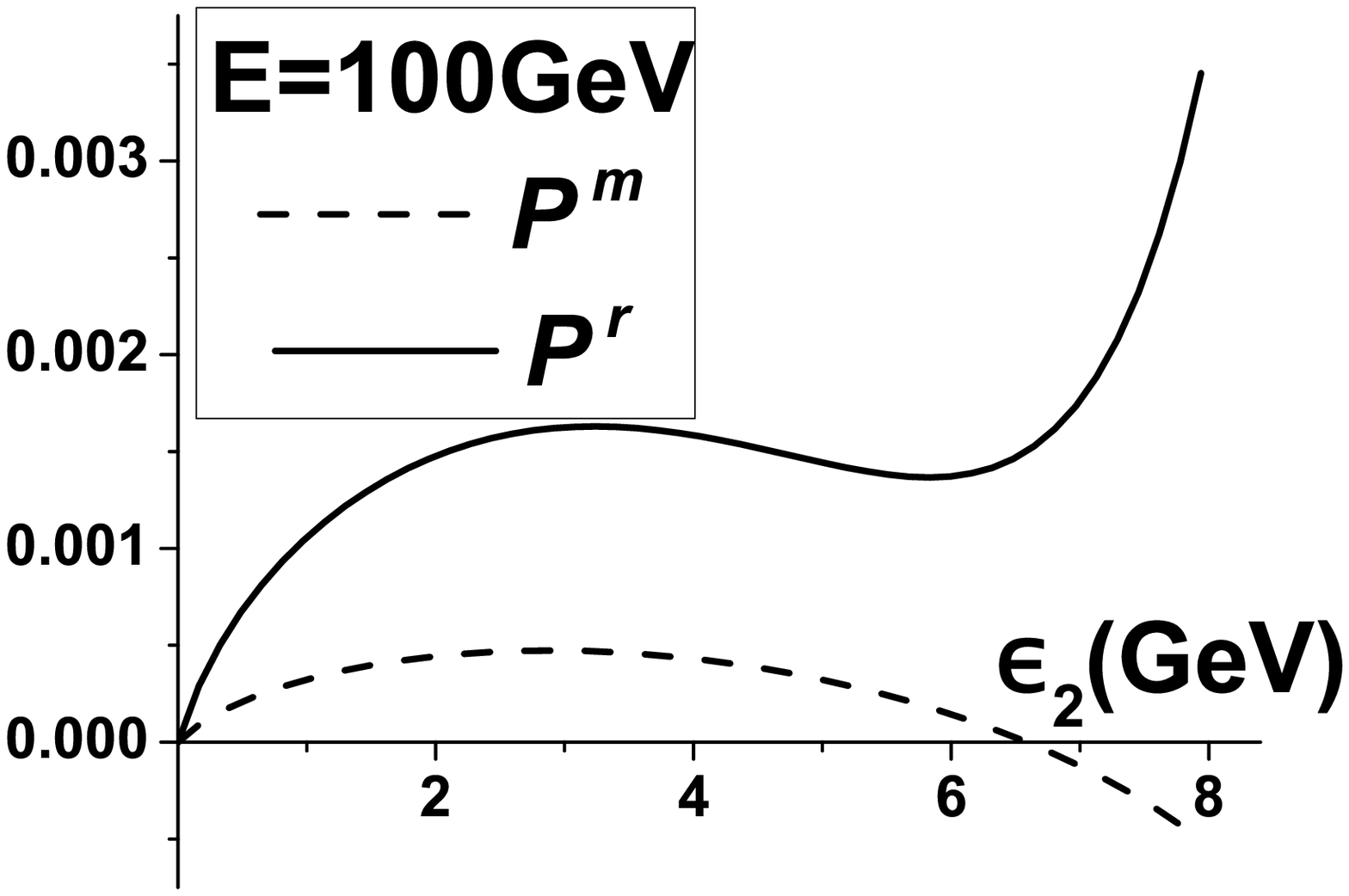}
\includegraphics[width=0.4\textwidth]{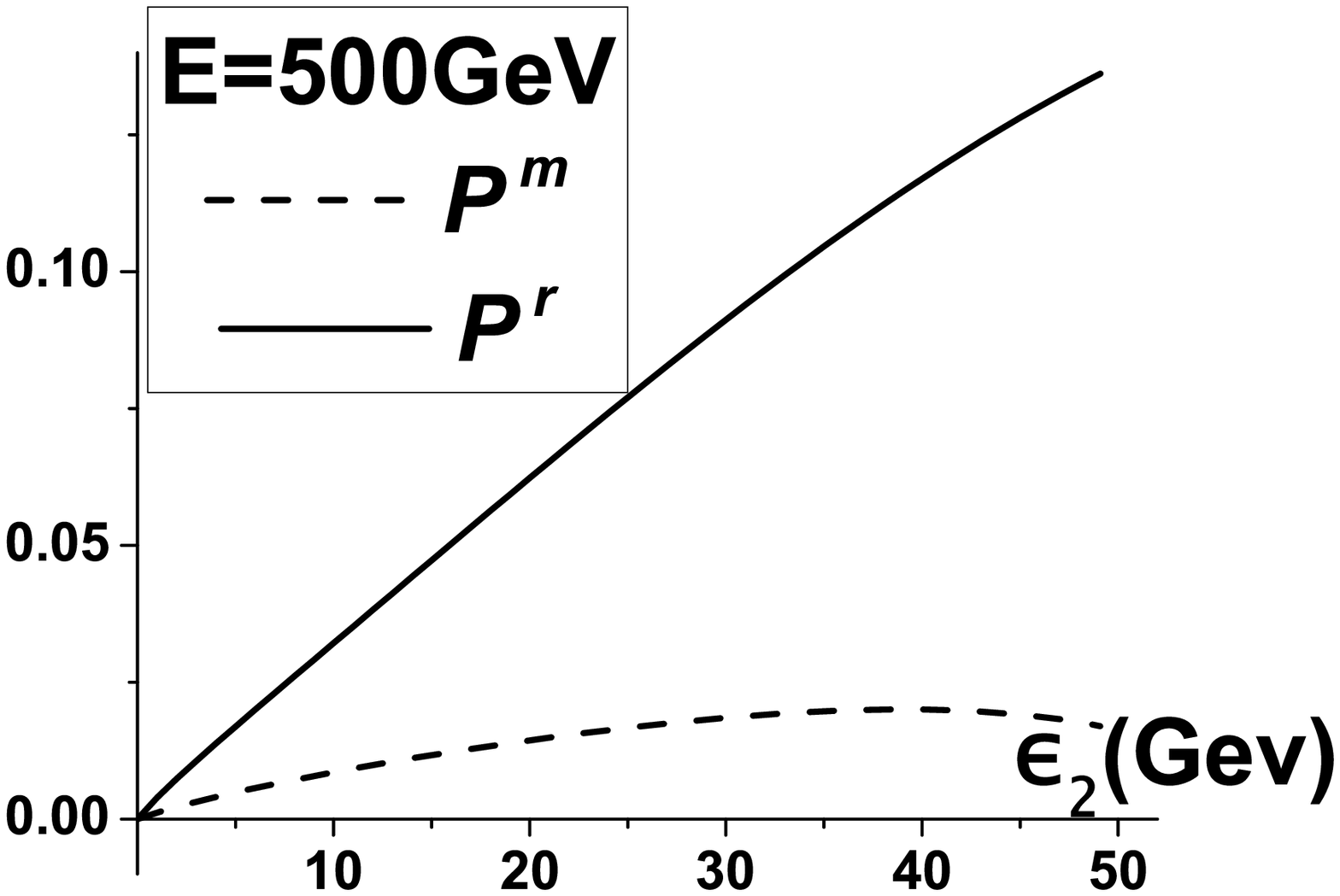}
 \parbox[t]{0.9\textwidth}{\caption{The sensitivity of the total model-independent radiative correction (in percent) to the choice of the form factor parametrization (Eq~(\ref{eq:62})).}
 \label{RCparametrization}}
\end{figure}

In Fig.\ref {RCparametrization} we illustrate the sensitivity of the total radiative correction to the parametrization of the form factors in terms of the ratios
\be\label{eq:62}
P^i=\frac{1+\delta_{\rm tot}^i}{1+\delta_{\rm tot}} -1\,, \ \ i=r,\,m\,,
\ee
where $\delta_{\rm tot}$ is the total correction for standard dipole fit.
We see that, in the considered conditions,  the deviation of these quantities from unity is very small and conclude that influence of the parameterizations
of the form factors on the radiative correction is much smaller than this influence on the Born cross section. Moreover, the $r$ and $m$ parameterizations decrease the Born cross section relative to the $d$ one,
as it follows from Fig.\ref{parameterization}, whereas for the radiative correction we have just opposite effect.

\section{Conclusion}

In present paper we investigated the recoil-electron energy
distribution in elastic proton-electron scattering in coincidence
experimental setup, taking the model-independent QED radiative
corrections into account. The detection of the recoil electron in
this process, with energies from a few MeV up to a few tens GeV,
will allow to receive the small-$Q^2$ data, at 10$^{-5}$\,GeV$^2$
$\leq Q^2\leq$ 3$\cdot$10$^{-2}$\,GeV$^2.$ Such data, being combined
with the existing and planning in the future experiments with the
electron beams, will help to perform more precise analysis of the
small-$Q^2$ behavior of the proton electromagnetic form factors. It
allows to obtain meaningful extrapolation to the static point and to
extract the proton charge radius. As noted in the recent review
\cite{Hill:2017puzz}, it is interesting to extract the proton charge
radius entirely from low-$Q^2$ data. High precision measurements, in
the inverse kinematics, allow to accumulate a lot of such data.

To cover the above mentioned interval of the $Q^2$-values, it is
desirable to use the proton beams with large enough energies, of the
order of a few hundreds GeV. At very small $Q^2,$ the sensitivity of
the differential cross section to the form factors parameterizations
is practically absent, but at $Q^2\approx$\,2$\cdot$10$^{-3}$\,GeV$^2$ it
has become noticeable and reaches several percent at
$Q^2\approx$\,3$\cdot$10$^{-2}$\,GeV$^2$ (see Fig.\,7). As follows from the
relations (\ref{eq:57}), the sensitivity to the value of the proton
charge radius also increases essentially with the growth of the
proton beam energy.

The effect, caused by the changing of the form factors
parametrization in the small-$Q^2$ region, is rather small.
Therefore, the accuracy of the measurement has to be high enough. In
Ref.\,\cite{Vorobyev:2016}, it is noted that in planning experiment
at the Mainz Microtron, with detection of the recoil proton, the
measurement precision has to be at the level of 0.2\%. To
discriminate between different form factor parameterizations the
accuracy, in the inverse kinematics experiments, must be the same,
possibly somewhat less with growth of $Q^2$ and the proton beam
energy. At such conditions the radiative corrections have to be
under control.

We account for the first order QED corrections caused by the vacuum
polarization and the radiation of the real and virtual photons by
the initial and final electrons, paying special attention to the
calculation of the hard photon emission contribution when the final
proton and electron energies are determined. This hard radiation
takes place due to the imprecision in the measurement of the proton
(electron) energy, $\Delta E\, \ (\Delta \varepsilon_2)$. In our
calculations we follow Ref.\, \cite{Kahane:1964zz} in choice of the
coordinate system and the angular integration method. We derive
analytical (although very cumbersome) expressions for the functions
$C_1(\omega)$ and $C_2(\omega),$ defined by Eqs.(\ref{eq:47}). The
cancelation of the auxiliary infrared parameter $\bar{\omega}$ in
the sum of the soft and hard corrections is performed analytically
and the rest $\omega-$integration in (\ref{eq:46}) is done
numerically.

The increase of the parameter $\Delta \epsilon_2$ leads to the small
decrease of the hard photon correction. The magnitude of this
decrease is about 0.01 (0.025) $\%$ at E=100 (500) GeV. Contrary,
the increase of the parameter $\Delta E_2$ increases the hard photon
correction by $\approx $ 0.2 (1)$\%$ at E=100 (500) GeV (see
Fig.\,8). Such different behaviour of this correction can be
explained, on the qualitative level, by Eq.\,(\ref{eq:60}).

As usually, there is a strong cancellation between the positive hard
correction and negative virtual and soft ones, as it is seen in
Fig.\,9. Despite the fact that the absolute values of these
corrections separately reach 20\,$-$40$\%$, their sum $|\delta_{\rm
tot}|$ does not exceed 6$\%$ at E=100 GeV and 4$\%$ at E=500 GeV for
the values $\Delta E_2 =0.02\,E\,, \ \Delta
\varepsilon_2=0.3\,\varepsilon_2$ used in calculations. The total
correction shows the very weak dependence on the form factors
parametrization (see. Fig.\,10) in the considered region. At the
lower values of $Q^2,$ which correspond to the lower values of the
recoil electron energy $\varepsilon_2,$ the total correction
$\delta_{\rm tot}$ is positive and changes sign when $Q^2$
increases. Such behaviour of $\delta_{\rm tot}$ is similar to the
one found in Ref.\, \cite{Kahane:1964zz} and confirmed in paper
\cite{Bardin:1970yn} for the case of the pion electron scattering.

In the papers \cite{Kahane:1964zz,Bardin:1970yn}, the authors
calculated also the model-dependent part of the radiative
corrections, using the point-like scalar electrodynamics to describe
the interaction of the charged pion with a photon.

In our case, the effect of the model-dependent corrections, when
virtual and real photons are emitted by the initial and final
protons, can be roughly evaluated in the approximation of the
structureless proton with the minimal proton-photon interaction
($F_1(q^2)=1\,, \ F_2(q^2)=0$), under the condition $Q^2/M^2\ll 1.$
The account for the proton structure can not change estimation
essentially. To derive the virtual and soft corrections in this
approximation, it is enough to write, in the expression ($\delta_0
+ \bar{\delta}$) defined in Eq.\,(\ref{eq:35}) (where $\bar{\delta}$
is taken without the term proportional to $M^2$), $\varepsilon_2$
and $k_2$ in terms of $Q^2$ and $m$, after that to change $m\to M$
and then to use the condition $Q^2/M^2\ll 1.$ Such procedure gives
their sum as
\be\label{eq:63}
\frac{\alpha}{\pi}\frac{Q^2}{M^2}\bigg(\frac{85}{36}+
\frac{4}{3}\ln\frac{M^2}{Q^2}+\frac{2}{3}\ln\frac{\bar{\omega}}{M}-2\ln{2}\bigg)\,.
\ee

The largest negative term with unphysical parameter $\bar{\omega}$
has to be cancelled by the hard photon contribution. The rest terms,
in (\ref{eq:63}) at $Q^2$=3$\cdot$10$^{-2}$\,GeV$^2$, do not exceed
0.5$\cdot$10$^{-3}$ that is a few times smaller than the required
measurement accuracy. In this approximation, the vertex correction
modifies also the Born value of the proton anomalous magnetic moment
and that leads to
$$G_M(q^2)\to
G_M(q^2)+\frac{\alpha}{2\,\pi}\bigg(1+\frac{q^2}{6\,M^2}\bigg)\,.$$
Taking $\mu_p=2.793$, we evaluate that the correction to the Born
value of $G_M^2$ is no more than 0.8$\cdot$10$^{-3}.$
The up-down interference, that is part of model-dependent QED corrections, which include the two-photon exchange and the interference of the proton and electron
radiation, has to be suppressed in considered kinematics at least by the factor
$$\frac{\alpha}{\pi}\,\frac{Q^2}{2m\,E}\,\ln\frac{2m\,E}{Q^2}\,,$$
that is no more than 0.1$\%.$ The analysis of the two-photon-exchange in the elastic $e^-$-p
scattering at small-$Q^2$, performed in \cite{Gorchtein:2014gm} for the more realistic model, confirms such suppression.


Thus, we conclude that the model-independent part of the radiative
corrections are under control and, if necessary, it can be
calculated with a more high accuracy. We believe that the
uncertainty due to the model-dependent part in the region $Q^2\ll
M^2$ is small and can not affect the experimental cross sections
measured even with 0.2$\%$ accuracy.


\section{Acknowledgments}
This work was partially supported by the Ministry of Education and
Science of Ukraine (projects no. 0115U000474 and no. 0117U004866).
The research is conducted in the scope of the IDEATE International Associated Laboratory (LIA).
\begin{appendix}

\section*{Appendix}
 \setcounter{equation}{0}
\def\theequation{A.\arabic{equation}}
 \label{sec:}

Since we use the expressions for the proton form factor expanded into series up to term of the order of  $q_1^4$ and the function $W_1(q_1^2)$ is proportional to $q_1^2$
we have to calculate the integrals (45) over the variables $\varphi$ and $y$ with the following integrands
 $S_1/(q_1^2\,|\vec{p}-\vec{k}|)\,, S_1/|\vec{p}-\vec{k}|,\\ S_2/(q_1^4\,|\vec{p}-\vec{k}|)\,,
\ S_2/(q_1^2\,|\vec{p}-\vec{k}|)\,, \ S_2/|\vec{p}-\vec{k}|\,.$

Let us introduce the following notation for integrals
\be\label{a.1}
J_{i\beta}=\int\limits_{y_-}^{y_+}d\,y\int\limits_0^{2\pi}d\,\varphi\,\frac{S_i}{(q_1^2)^\beta|\vec{p}-\vec{k}|}, \ \ I_{i\beta}=\int\limits_{y_-}^{\bar{y}}d\,y\int\limits_0^{2\pi}d\,\varphi\,\frac{S_i}{(q_1^2)^\beta|\vec{p}-\vec{k}|}\,, \ i=1,\,2, \ \beta=0,\,1,\,2;
\ee

$$\ d_{\pm}=\omega+\epsilon_2\pm m\,, \ k_h=\sqrt{d_+\,d_-}\,, \ \epsilon_{\pm}=\epsilon_2\pm m\,, \ \ Z=m^2+M^2+2mE\,,$$
$$E_2=m+E-\omega-\epsilon_2,\,  \left|
   \overset{\rightharpoonup }{p_2}\right|=\sqrt{(m+E-\omega-\epsilon_2)^2-M^2}.$$
Then we have
 \be
  J_{22}=\frac{2 \pi }{\omega ^2} \Big[C_{22}+\frac{\overline{C}_{22}}{m\omega}\,
 L^{(\omega)}+\frac{2\widetilde{C}_{22}}{m\,k_h}L^{(h)}\Big]\,, \
   \ee
$$ L^{(\omega)}=\ln\Big(1-\frac{2\omega\,d_-}{m\epsilon_-}\Big)\,, \ \ L^{(h)}=\ln\frac{\omega+\epsilon_2+k_h}{m}\,, $$
 $$C_{22}=2+\frac{\epsilon_+ [7 m^3+2 E \left(4 m-3 \epsilon _2\right) m+E^2
   \left(m-6 \epsilon _2\right)+\left(M^2-m^2\right) \epsilon _2]}{2\,m^3\,d_+}-$$
 $$\frac{6m^3\epsilon_- -
   \left(8 M^2+\left| \overset{\rightharpoonup }{k_2}\right|^2\right) m^2+6 E \left(4
   m^2-\epsilon _2 \epsilon _+\right) m-M^2 \epsilon _2 \left(3 m+7 \epsilon _2\right)-E^2 \left(\left| \overset{\rightharpoonup }{k_2}\right|^2+5 \epsilon _2 \epsilon_+\right)}{2\,m^3\,d_-}+$$
 $$\frac{4 \epsilon _2\,M^2 [m \left(\epsilon _2-\omega \right)-2 \omega  \epsilon _2]}{m^3[m\omega+d_-(2\omega-m)]}
 +\frac{\left(8 m^2-5 \epsilon _2 m-\epsilon _2^2\right) M^2+2 m E \left(9 m-2 \epsilon _2\right) \epsilon _-+3 E^2 \left(4 m^2-\epsilon _2
   \epsilon _+\right)}{m^2\,d_-^2}$$
 $$-\frac{3 (m+E)^2\epsilon_+^2}{m^2\,d_+^2}+\frac{2\epsilon_-[2 E m^2+E^2 \left(2 \epsilon _2-9
   m\right)+M^2 \left(2 \epsilon _2-5 m\right)]}{m\,d_-^3}-\frac{2m\epsilon_-(3M^2+2E^2)}{d_-^4}\,,$$
$$\overline{C}_{22}=\frac{2[(\epsilon_2-2m)M^2-2mE(E+2\epsilon_-)]}{d_-} -\frac{\epsilon_-[3 M^2 \left(\epsilon _2-3 m\right)-2 m E \left(-5 m+4 E+3 \epsilon _2\right)]}{d_-^2}+$$
$$\frac{\epsilon_-[\left(12 m^2-9 \epsilon _2 m+\epsilon _2^2\right) M^2+2 m E^2 \left(5 m-3 \epsilon _2\right)+2 m E \epsilon _- \left(3
   m-\epsilon _2\right)]}{d_-^3}+$$
$$\frac{2m\,\epsilon_-^2[\left(M^2+E^2\right)
   \epsilon _2-m \left(4 M^2+3 E^2-m E\right)]}{d_-^4}-\frac{\epsilon_-^2m^3(3M^2+2E^2)}{d_-^5}+4m\,E\,,$$

$$\widetilde{C}_{22}=\frac{\epsilon_-\left(5 m-\epsilon _2\right) m^2+2m\,E \left(2 m-\epsilon _2\right) \left(3 m-\epsilon _2\right)+p^2 \epsilon _2\epsilon_+
  +E^2 \left(5 m^2+3 \epsilon _2^2\right)}{4m\,d_-}+$$
$$\frac{\epsilon_+[9m^3+2m\,E(6m-\epsilon_2)+m^2\epsilon_2-3E^2\epsilon_--p^2\epsilon_2]}{4m\,d_+}-\frac{3(m+E)^2\epsilon_+^2}{2\,d_+^2}-$$
$$\frac{10m^2E\epsilon_- +E^2(2m-\epsilon_2)(3m-\epsilon_2)+M^2\epsilon_-(2m+\epsilon_2)}{2\,d_-^2}-\frac{5m^2E^2\epsilon_-}{d_-^3}-2m^2-2mE-M^2\,,$$

\be
J_{21}=\frac{8 \pi }{\omega ^2} \Big[C_{21}+\frac{\overline{C}_{21}M^2}{2\omega}\,L^{(\omega)}+\frac{\widetilde{C}_{21}}{k_h}L^{(h)}\Big]\,,
\ee
$$C_{21}=\frac{2\omega^2(3E+2m-\omega)}{3m}-\frac{\omega(2E^2+2mE+m^2)+\epsilon_2(M^2+m^2)+2(m^3+m^2E-\epsilon_2E^2)}{m}+$$
$$\frac{\epsilon_+[2m^2(7E+5m)+\epsilon_2(2m^2+M^2)+E^2(4m-3\epsilon_2)]}{2m\,d_+}-\frac{3(E+m)^2\epsilon_+^2}{d_+^2}-$$
$$\frac{\epsilon_-(6m^2E+4mM^2+p^2\epsilon_2)-2mE^2\epsilon_2}{2m\,d_-}+\frac{m\epsilon_-(3E^2-M^2)}{d_-^2}\,,$$

$$\overline{C}_{21}=-2(\omega+m-\epsilon_2)+\frac{\epsilon_-(5m-3\epsilon_2)}{d_-}-\frac{\epsilon_-^2(3m-\epsilon_2)}{d_-^2}-\frac{m^2\,\epsilon_-^2}{d_-^3}\,,$$

$$\widetilde{C}_{21}=\omega ( m^2+Z)+\epsilon _2(-3 m^2+Z)-m [3 m^2+2 E (E+3 m)]+$$
     $$\frac{\epsilon_- \left(6 E m^2+M^2 \epsilon _2\right)+E^2 \left(2 m^2-5 m \epsilon _2+\epsilon
   _2^2\right)}{2 d_-}-\frac{3 E^2 m^2 \epsilon_-}{d_-^2}+$$
   $$+\frac{\epsilon_+}{2 d_+} \big[2 m (E+m) \left(3 E+6 m+2 \epsilon
   _2\right)-\epsilon _2 p^2 \big]-\frac{3 m (E+m)^2 \epsilon_+^2}{d_+^2}\,.$$

\be
J_{20}=\frac{16 \pi }{\omega ^2}\Big[C_{20}
+\frac{\widetilde{C}_{20}m}{k_h}L^{(h)}\Big]\,,
\ee

$$C_{20}=-\frac{4\omega^5}{5\,m}+\frac{2 \omega ^4 \left(20 E+19 m-10 \epsilon _2\right)}{15 m}-\frac{\omega ^3}{3 m} \Big[2 \left(4 E^2+6 m^2+9 E m\right)-8\epsilon _2 (E+m)-M^2\Big]+$$
$$\frac{\omega ^2}{3 m} \Big[11 m^3-\epsilon _2 \left(8 m^2+3
   M^2\right)+2 m \left(5 E^2-M^2\right)+2 E m \left(11 m-5 \epsilon _2\right)\Big]+$$
$$\omega[m^3+2m(M^2-2E^2)+\epsilon_2(2m^2+4mE+M^2)]-\frac{E^2 m^2 \epsilon _-}{d_-}-\frac{6 m^2 (E+m)^2 \epsilon _+^2}{d_+^2}+$$
$$+\frac{m\,\epsilon_+ \left[m \left(7 E^2+13 m^2\right)+\epsilon _2 \left(5 m^2+M^2\right)+2
   E m \left(10 m+3 \epsilon _2\right)\right]}{d_+}\,,$$

$$\widetilde{C}_{20}= -\omega ^2 (m^2+Z)+\omega[m^3+2m(E+m)(E+2m)-2\epsilon _2(2mE+M^2)]-m^3(E-2\epsilon _2)-$$
$$m^2(4E+9m)(E+\epsilon_+)-2mE\,\left| \overset{\rightharpoonup }{k_2}\right|^2+\epsilon_2 M^2(m-2\epsilon_2)+2m \epsilon_2\,E(E-\epsilon_+)+ \frac{m^2E^2\epsilon_-}{d_-}+$$
$$\frac{6 m^2 (E+m)^2 \epsilon _+^2}{d_+^2}+\frac{m\epsilon_+[\epsilon_2(2E^2+10mE+7m^2+M^2)+3m(E+m)(3E+5M)]}{d_+}\,.$$

\be
J_{11}=\frac{16 \pi}{\omega ^2}\Big[C_{11}-
\frac{\overline{C}_{11}L^{(\omega)}}{2\omega d_ -^3}+
\frac{\widetilde{C}_{11}}{k_h}L^{(h)}\Big]\,,
\ee
$$C_{11}= -\frac{3 m^4-m^3 \left(\omega +2 \epsilon _2\right)-m^2 \left(2 \omega ^2+5 \omega  \epsilon _2+3 \epsilon
   _2^2\right)+2 m \epsilon _2 \left(\omega +\epsilon _2\right){}^2+\omega  \left(\omega +\epsilon _2\right){}^3}{m d_-^2\,d_+}\,,$$

$$\overline{C}_{11}=2\omega^4+4\omega^3\epsilon_-  +\omega^2(5m^2-8m\epsilon_2+3\epsilon_2^2)-\omega\epsilon_-^2(3m-\epsilon_2) +m^2\epsilon_-^2\,,$$
$$\widetilde{C}_{11}=-\frac{3 m^2\epsilon_-}{2 d_-{}^2}
    +\frac{7 m^2-13 m \epsilon _2+4 \epsilon _2^2}{4 d_-}-\frac{3 m \epsilon_+}{4 d_+}+d_+\,.$$

\be
J_{10}=\frac{16 \pi}{\omega ^2}\Big[C_{10}-
\frac{m\,\widetilde{C}_{10}}{k_h^3}L^{(h)}\Big]\,,
\ee
$$C_{10}=-m^2\big[\frac{\epsilon_-}{d_-}+\frac{3\epsilon_+}{d_+}\big]-\frac{2\omega^2(d_++2\epsilon_2)}{3\,m}+2\omega\epsilon_++4m\epsilon_-\,,$$
$$\widetilde{C}_{10}=(\epsilon_2+\omega)^2\,[2\,\left| \overset{\rightharpoonup }{k_2}\right|^2+m(\omega-4\epsilon_2)+\omega^2+2\omega\epsilon_2]+m^2[\omega\epsilon_++4m\,\epsilon_2]\,.$$

$$I_{22}=\frac{2\pi}{\omega^2}\Big[D_{22}+\frac{D_{22}^{L_1}}{m\,k_h}L_1+\frac{D_{22}^{L_2}}{m\,\omega}L_2+\frac{1}{m\omega K_2}D_{22}^2+
\frac{\omega K_1}{m^2k_h\,d_-}\overline{D}_{22}^1+\frac{K_2}{m^2\omega\,d_-^3}\overline{D}_{22}^2 +$$
  \be
\frac{1}{m\omega(2mp^2-d_-Z)}\Big(\frac{2mE^2-d_-(M^2+2mE)}{k_h\,K_1}D_{22}^1+p\,D_{22}^p\Big)\Big]\,,
\ee

$$K_1=\sqrt{\frac{m^2(2mp^2-d_-Z)}{d_+^2\,d_-}+\frac{1}{\omega^2}\big(\left| \overset{\rightharpoonup }{k_2}\right|  \left|\overset{\rightharpoonup }{p_2}\right| -\frac{(E_2-m)(\left| \overset{\rightharpoonup }{k_2}\right|^2+\omega\epsilon_2)}{d_+}\Big)^2}\,,$$
$$K_2=\sqrt{\big(\left| \overset{\rightharpoonup }{k_2}\right|  \left|\overset{\rightharpoonup }{p_2}\right| d_--\epsilon_-(\epsilon_2+\omega)E_2\big)^2+m\epsilon_-M^2(2\omega d_--m\epsilon_-)}\,,$$
$$L_1=\ln\frac{\omega k_h^2K_1+d_-[\left| \overset{\rightharpoonup }{k_2}\right|  \left|\overset{\rightharpoonup }{p_2}\right| d_+-(E_2-m)(\left| \overset{\rightharpoonup }{k_2}\right|^2+\omega\epsilon_2)]}{\omega(\omega+\epsilon_2-k_h)[pk_h+(E+m)d_-]}\,,$$
$$L_2=\ln\frac{K_2+\epsilon_-E_2(\omega+\epsilon_2)-\left| \overset{\rightharpoonup }{k_2}\right|  \left|\overset{\rightharpoonup }{p_2}\right| d_-}{m\epsilon_-(E+p)}\,,$$

$$D_{22}=1+\frac{2\epsilon_2M^2(m(\epsilon_2-\omega)-2\omega\epsilon_2)}{m^3(2\omega d_--m\epsilon_-)}+\frac{\epsilon_+}{4m^3d_+}\Big[E^2(m-6\epsilon_2)
+2mE(4m-3\epsilon_2)+\epsilon_2(M^2-m^2)+7m^3\Big]-$$
$$\frac{1}{4m^3d_-}\big[\epsilon_+E^2(m-6\epsilon_2)+6mE(4m^2-\epsilon_2\epsilon_+)-\epsilon_2M^2(3m+7\epsilon_2)-m^2(8M^2+\epsilon_2^2)+
m^3(6\epsilon_2-5m)\big]-$$
$$\frac{3\epsilon_+^2(E+m)^2}{2m^2d_+^2}+\frac{1}{2m^2d_-^2}\big[3E^2(4m^2-\epsilon_2\epsilon_+)+2m\epsilon_-E(9m-2\epsilon_2)+
M^2(8m^2-5m\epsilon_2-\epsilon_2^2)\big]+$$
$$\frac{\epsilon_-}{md_-^3}\big[E^2(2\epsilon_2-9m)+2m^2E+M^2(2\epsilon_2-5m)\big]-\frac{m\epsilon_-(3M^2+2E^2)}{d_-^4}\,,$$

$$D_{22}^{L_1}=-2m(E+m)-M^2+\frac{\epsilon_+}{4md_+}[9m^3+2m\,E(6m-\epsilon_2)+\epsilon_2(m^2-p^2)-3\epsilon_-E^2]+\frac{1}{4md_-}[p^2\epsilon_2\epsilon_+\,+$$
$$12m^3E+E^2(5m^2+3\epsilon_2^2)+m(5m-\epsilon_2)(m\epsilon_--2\epsilon_2E)]-\frac{3\epsilon_+^2(E+m)^2}{2d_+^2}+$$
$$\frac{1}{2d_-^2}[-m\epsilon_-(10mE+M^2)-M^2\left| \overset{\rightharpoonup }{k_2}\right|^2-E^2(6m^2-5m\epsilon_2+\epsilon_2^2)]+\frac{5m^2E^2\epsilon_-}{d_-^3}\,,$$

$$D_{22}^{L_2}=4mE+\frac{2}{d_-}[\epsilon_2M^2-4m\epsilon_-E-2m(E^2+M^2)]+\frac{\epsilon_-}{d_-^2}[2mE(4E+3\epsilon_2-5m)-3M^2(\epsilon_2-3m)]+$$
$$\frac{\epsilon_-}{d_-^3}[M^2(12m^2-9m\epsilon_2+\epsilon_2^2)+2mE^2(5m-3\epsilon_2)+2m\epsilon_-E(3m-\epsilon_2)]+$$
$$\frac{2m\epsilon_-^2}{d_-^4}[m^2E+M^2(\epsilon_2-4m)+E^2(\epsilon_2-3m)]-\frac{m^3\epsilon_-^2(3M^2+2E^2)}{d_-^5}\,,$$

$$\overline{D}_{22}^1=\left| \overset{\rightharpoonup }{k_2}\right|  \left|\overset{\rightharpoonup }{p_2}\right| +3m^2+\epsilon_2(d_+-E)-\frac{3m\epsilon_+(m+E)}{d_+}-\frac{4m^2E}{d_-}\,, \ \
\overline{D}_{22}^2=\omega^2\epsilon_++\omega[\left| \overset{\rightharpoonup }{k_2}\right|  \left|\overset{\rightharpoonup }{p_2}\right| +\epsilon_2^2-$$
$$\epsilon_2(m+E)-3mE]+m[m(2m-7E)+\epsilon_2(3E-2m)]+\frac{\epsilon_-mE(7m-3\epsilon_2)}{d_-}\,,$$

$$D_{22}^1=\left| \overset{\rightharpoonup }{k_2}\right|^2\,Z+m^2\epsilon_2(E+m)+\frac{1}{8d_-}\big\{4\left| \overset{\rightharpoonup }{k_2}\right|  \left|\overset{\rightharpoonup }{p_2}\right| [2\epsilon_2(M^2+mE)+\epsilon_+(m^2-E^2)]-$$
$$\epsilon_+E^3(3m-5\epsilon_2)-mE^2[6m\epsilon_2-5(5m^2-3\epsilon_2)]+\epsilon_+E[4m^2\epsilon_--M^2(9\epsilon_2-7m)]-$$
$$m[8m^3\epsilon_--\epsilon_+M^2(17m-15\epsilon_2)]\big\}+\frac{\epsilon_+(E+m)}{8d_+}\big[4\left| \overset{\rightharpoonup }{k_2}\right|  \left|\overset{\rightharpoonup }{p_2}\right| (E+m)-8m^2\epsilon_2+$$
$$4mE(m-3\epsilon_2)+E^2(3m-5\epsilon_2)+\epsilon_+M^2\big]+\frac{1}{4d_-^2}\big\{4m\left| \overset{\rightharpoonup }{k_2}\right|  \left|\overset{\rightharpoonup }{p_2}\right| [\epsilon_-(M^2+mE)-\epsilon_2p^2]+$$
$$m\epsilon_-[p^2E(5m+3\epsilon_2)-mE^2(7m+\epsilon_2)+4m^2E\epsilon_-+m M^2(3m+5\epsilon_2)]\big\}-\frac{m^2p^2\epsilon_-}{2d_-^3}[2\left| \overset{\rightharpoonup }{k_2}\right|  \left|\overset{\rightharpoonup }{p_2}\right| +\epsilon_-(E+m)]\,,$$


$$D_{22}^2=\left| \overset{\rightharpoonup }{k_2}\right|  \left|\overset{\rightharpoonup }{p_2}\right| \Big[d_-[2m(E+2\epsilon_-)-M^2]-2mE^2+3\epsilon_-(2m^2-4mE+M^2)-\epsilon_2\,M^2\Big]+$$
$$\omega^2[2m\epsilon_-(E+2\epsilon_2)-M^2\epsilon_+]+\frac{5m^4\epsilon_-^3p^2E}{d_-^5}+$$
$$\omega\big[mE(M^2-4\epsilon_-E)-2m\epsilon_2^2(5m+6E-4\epsilon_2)-2m^3(E-3m)+\epsilon_2[E(14m^2+M^2)-4m^3]\big]+\frac{\epsilon_-}{2}\big[2m^2(5M^2-$$
$$8E^2)+
2mE(25m^2-3M^2+2E^2)+2\epsilon_2E(16mE-18m^2-7m\epsilon_2-M^2)+\epsilon_2\epsilon_-(8m\epsilon_2-M^2)\big]-$$
$$\frac{\epsilon_2M^2}{2m(2\omega d_--m\epsilon_-)}\big\{\left| \overset{\rightharpoonup }{k_2}\right|^2[2E(2\omega\epsilon_2-md_-)-m\epsilon_-(m-2\omega)]+2\left| \overset{\rightharpoonup }{k_2}\right|  \left|\overset{\rightharpoonup }{p_2}\right| \epsilon_-[m\omega-\epsilon_2
(m-2\omega)]\big\}+$$
$$\frac{\epsilon_-}{md_-}\big\{\left| \overset{\rightharpoonup }{k_2}\right|  \left|\overset{\rightharpoonup }{p_2}\right| [2m^2(3M^2+4E^2)+m^2E(7\epsilon_2-23m)-\epsilon_-(2m^3-\epsilon_2M^2)]
+2m^2p^2[\epsilon_-(20m-7\epsilon_2)+$$
$$E(3m-4\epsilon_2)]+E[2m^3(17m^2-3M^2)-m^2\epsilon_2(57m^2+5M^2)+\epsilon_2^2\big(2m(12m^2+M^2)+\epsilon_2(M^2-m^2)\big)]+$$
$$m\epsilon_-[M^2(30m^2-17m\epsilon_2+\epsilon_2^2)-3m^3\epsilon_-]\big\}+\frac{\epsilon_-}{d_-^2}\big\{\left| \overset{\rightharpoonup }{k_2}\right|  \left|\overset{\rightharpoonup }{p_2}\right| m\big[E^2(17m-7\epsilon_2)+2M^2(4m-3\epsilon_2)
-$$
$$\epsilon_-\big(m^2+E(\epsilon_2-17m)\big)\big]-\epsilon_-\big[mE^3(17m-7\epsilon_2)-2mE^2(25m^2-20m\epsilon_2+\epsilon_2^2)
+E\big(m^2(22m^2-23m\epsilon_2+$$
$$\epsilon_2-23M^2)+M^2\epsilon_2(2m+\epsilon_2)\big)+m^2(m^2\epsilon_-+9mM^2+\epsilon_2M^2)\big]\big\}
+\frac{m\epsilon_-^2}{d_-^3}\big\{\left| \overset{\rightharpoonup }{k_2}\right|  \left|\overset{\rightharpoonup }{p_2}\right| [E^2(\epsilon_2-15m)+$$
$$8m^2E+M^2(\epsilon_2-7m)]-mE^3(22m-19\epsilon_2)-\epsilon_2^2p^2E-mE[M^2(14\epsilon_2-27m)-7m^2\epsilon_-]+$$
$$m\epsilon_-[3mM^2+E^2(2\epsilon_2-35m)]\big\}-\frac{m^2\epsilon_-^2}{d_-^4}\big\{m\left| \overset{\rightharpoonup }{k_2}\right|  \left|\overset{\rightharpoonup }{p_2}\right| (7E^2+3M^2)+\epsilon_-[m^2(11E^2-M^2)-p^2E(16m-\epsilon_2)]\big\}\,,$$

$$D_{22}^p=-2\,Z[\omega(2m-\epsilon_2)+\epsilon_2^2]-11m^4+18m^2p^2+5mE(M^2-3m^2)+2\epsilon_2[M^2(6m-E)+$$
$$m(4m^2+7mE-4E^2)]+\frac{\epsilon_+^2(m+E)}{4md_+}\big[E^2(7m-6\epsilon_2)+2mE(12m-\epsilon_2)+m(15m^2+2M^2)+\epsilon_2(m^2+3M^2)\big]-$$
$$\frac{1}{4md_-}\big\{E^3[m^2(47m-8\epsilon_2)-\epsilon_2^2(5m+6\epsilon_2)]-mE^2[m^2(241m-302\epsilon_2)+\epsilon_2^2(49m+8\epsilon_2)]
+E[m^4(75m-$$
$$167\epsilon_2)+m^3(109\epsilon_2^2-118M^2)+m^2\epsilon_2(91M^2-17\epsilon_2^2)+\epsilon_2^2M^2(3\epsilon_2-8m)]-m\epsilon_-[m^3(79m-54\epsilon_2)+$$
$$7m^2(26M^2+\epsilon_2^2)+\epsilon_2M^2(5\epsilon_2-79m)]\big\}-\frac{3\epsilon_+^3(E+m)^3}{2d_+^2}
+\frac{\epsilon_-}{2d_-^2}\big\{E^3[5m(14m-3\epsilon_2)-3\epsilon_2^2]+$$
$$mE^2[m(123\epsilon_2-184m)-15\epsilon_2^2]-E[2m^3(3m+13\epsilon_2)+8m^2(16M^2-\epsilon_2^2)+\epsilon_2M^2(\epsilon_2-49m)]+$$
$$m[2m^3(13m-16\epsilon_2)+2m^2(52M^2+3\epsilon_2^2)+\epsilon_2M^2(13\epsilon_2-85m)]\big\}+\frac{m\epsilon_-}{d_-^3}\big\{E^3[4\epsilon_2^2
+5m(11m-7\epsilon_2)]+$$
$$mE^2(-61m^2+75m\epsilon_2-14\epsilon_2^2)-2E[m^3(5m-7\epsilon_2)+m^2(37M^2+2\epsilon_2^2)+\epsilon_2M^2(4\epsilon_2-29m)]-$$
$$2m\epsilon_-[2m^3+3M^2(5m-\epsilon_2)]\big\}+\frac{m^2\epsilon_-^2}{d_-^4}\big\{E^3(8\epsilon_2-36m)+18m^2E^2+E[5m^3+M^2(41m-8\epsilon_2)]-8m^2M^2\big\}-$$
$$-\frac{10m^4\epsilon_-^2p^2E}{d_-^5}\,,$$


$$I_{21}=\frac{4\pi}{\omega^2}\Big\{D_{21}+\frac{D_{21}^{L_1}}{k_h}L_1+\frac{M^2\,D_{21}^{L_2}}{\omega}L_2
-\frac{\omega\,K_1}{m\,k_h}\overline{D}_{21}^1+\frac{\overline{D}_{21}^2}{\omega\,d_-^2}K_2+$$
\be
\frac{1}{2mp^2-d_-Z}\Big[\frac{2mEE_2-M^2d_-}{k_h\,K_1}D_{21}^1+\frac{p}{3m\,\omega}D_{21}^p\Big]\Big\}\,,
\ee

$$D_{21}=-\frac{2\omega^3}{3m}+\frac{2\omega^2(3E+2m)}{3m}-\frac{\omega[m^2+2E(m+E)]}{m}-2m(m+E)+\frac{\epsilon_2(E^2+p^2-m^2)}{m}+$$
$$\frac{\epsilon_+}{2md_+}\big[2m^2(5m+7E)+E^2(4m-3\epsilon_2)+\epsilon_2(2m^2+M^2)\big]-\frac{1}{2m\,d_-}\big\{\epsilon_-[p^2\epsilon_2
-2m(E^2-2M^2-3mE)]-$$
$$2m^2E^2\big\}-\frac{3\epsilon_+^2(m+E)^2}{d_+^2}+\frac{m\epsilon_-(3E^2-M^2)}{d_-^2}+\frac{k_2+p_2}{3m\omega}\Big\{2\omega^2[-\omega+\epsilon_-+2(m+E)]+$$
$$\omega[p^2-3E^2+2\left| \overset{\rightharpoonup }{k_2}\right|  \left|\overset{\rightharpoonup }{p_2}\right| +2\epsilon_-(\epsilon_2-E)-2m(E-\epsilon_-)]+m[p^2-3E(2m+E)+2\left| \overset{\rightharpoonup }{k_2}\right|  \left|\overset{\rightharpoonup }{p_2}\right| +2\epsilon_-(\epsilon_2-E)]\Big\}\,,$$
$$D_{21}^{L_1}=(\omega+\epsilon_2)(M^2+2mE)-m[3m^2+2E(3m+E)]-2m^2(\epsilon_2-\omega)-\frac{3m^2\epsilon_-E^2}{d_-^2}-\frac{3m\epsilon_+^2(m+E)^2}{d_+^2}+$$
$$\frac{1}{2d_-}\big[\epsilon_-(6m^2E+\epsilon_2M^2)+E^2(\epsilon_2^2+2m^2-5m\epsilon_2)\big]+$$
$$\frac{\epsilon_+}{2d_+}\big[E^2(6m-\epsilon_2)+2mE(9m+2\epsilon_2)
+\epsilon_2(4m^2+M^2)+12m^3\big]\,,$$
$$D_{21}^{L_2}=-2(\omega-\epsilon_-)+\frac{\epsilon_-(5m-3\epsilon_2)}{d_-}-\frac{\epsilon_-^2(3m-\epsilon_2)}{d_-^2}-\frac{m^2\epsilon_-^2}{d_-^3}\,,$$
$$\overline{D}_{21}^1=\left| \overset{\rightharpoonup }{k_2}\right|  \left|\overset{\rightharpoonup }{p_2}\right| +\epsilon_2(\omega+\epsilon_2-E)+m(3m+\epsilon_2)-\frac{3m\epsilon_+(m+E)}{d_+}-\frac{4m^2E}{d_-}\,,$$
$$\overline{D}_{21}^2=\left| \overset{\rightharpoonup }{k_2}\right|  \left|\overset{\rightharpoonup }{p_2}\right| +\epsilon_-(\omega+\epsilon_2-E)-\frac{m\epsilon_-E}{d_-}\,,$$
$$D_{21}^1=-\frac{\left| \overset{\rightharpoonup }{k_2}\right|  \left|\overset{\rightharpoonup }{p_2}\right| -\epsilon_-(E_2-m)}{\omega}\Big\{\epsilon_2(m^2+M^2)+m[m(E+m)+2\epsilon_2E]-\frac{m\epsilon_+(m+E)^2}{d_+}-$$
$$\frac{m[\epsilon_2p^2-\epsilon_-(M^2+mE)]}{d_-}-\frac{m^2p^2\epsilon_-}{d_-^2}\Big\}-m[\epsilon_-(M^2+mE)+m\epsilon_2(m+E)]-$$
$$\frac{m\big[M^2[3m(E-3m)+\epsilon_2(7m-5E)]+E^2[m(13m-11\epsilon_2)+\epsilon_+E]\big]}{4d_-}+$$
$$\frac{m\epsilon_+(m+E)[p^2+4m(m+E)]}{4d_+}-\frac{m^2\epsilon_-p^2(3E-m)}{2d_-^2}\,,$$

$$D_{21}^p=-\omega^2\,Z(2E^2-6\epsilon_2E+M^2+6m^2)+\omega\big\{2p^2[8m^3+2mE(E-4\epsilon_2)-\epsilon_-(m^2+M^2)]+$$
$$3m^4(3m+2\epsilon_2)+3mE[m^2(5m+4\epsilon_2)+M^2(m-2\epsilon_2)]+3M^2[m^2(4m+\epsilon_2)-\epsilon_2M^2]\big\}-mp^2\big\{-4mE^2+$$
$$2E[m^2+2\epsilon_2(4m-3\epsilon_2)]+2[2m(m^2-M^2)+\epsilon_2(31m^2+M^2)]\big\}-3m\big\{2m^4(3m+4\epsilon_2)+m^3(5M^2+2\epsilon_2^2)-$$
$$mM^2(M^2-2\epsilon_2^2)+mE[2m^2(7m+10\epsilon_2)+4m\epsilon_2^2-M^2(m-8\epsilon_2)]+\epsilon_2M^2(20m^2+M^2)\big\}+$$
$$\frac{3m\epsilon_-}{2d_-}\big\{E^3(10m^2+7m\epsilon_2-5\epsilon_2^2)-mE^2[\epsilon_2^2+3m(4m-3\epsilon_2)]+E[\epsilon_-(5\epsilon_2M^2-2m^3)-$$
$$8m^2M^2]-m\epsilon_-M^2(6m-\epsilon_2)\big\}+\frac{3m\epsilon_+^2(m+E)}{2d_+}\big[E^2(10m-3\epsilon_2)+2mE(15m+2\epsilon_2)+$$
$$M^2(2m+3\epsilon_2)+2m^2(9m+2\epsilon_2)\big]-\frac{3m^3\epsilon_-^2E}{d_-^2}[m^2+M^2+E(m+E)]-\frac{9m^2\epsilon_+^3(m+E)^3}{d_+^2}+
\frac{6m^4\epsilon_-^2p^2E}{d_-^3}\,.$$

$$I_{20}=\frac{2\pi}{\omega}\Big\{-\frac{4}{15m}D_{20}+\frac{4m\,L_1}{k_h}D_{20}^{L_1}+\frac{4\omega\,k_h\,K_1}{d_+}\overline{D}_{20}^1-\frac{4\,p}{15m\,\omega}
\overline{D}_{20}^p+\frac{\bigl| \overset{\rightharpoonup }{k_2}\bigr| + \bigl|\overset{\rightharpoonup }{p_2}\bigr|}{15\,m\,\omega}\widetilde{D}_{20}-$$
\be
-\frac{4\,m}{\omega(2mp^2-d_-Z)}\Big[\frac{2mEE_2-M^2d_-}{k_h\,K_1}D_{20}^1+\frac{p}{2}D_{20}^p\Big]\Big\}
\ee

$$D_{20}=12\omega^5-2\omega^4(20E-19m-10\epsilon_2)+5\omega^3[8E^2+12m^2+18mE-8\epsilon_2(m+E)-M^2]-$$
$$-5\omega^2[11m^2(m+2E)+10mE(E-\epsilon_2)-M^2(2m+3\epsilon_2)-8m^2\epsilon_2]-$$
$$15m\omega[\epsilon_2(M^2+2m^2+4mE)-m(4E^2-m^2-2M^2)]+$$
$$15m^2[m(4E^2+7m^2+12mE+2M^2)+\epsilon_2(5m^2+2mE-M^2)]+\frac{15m^3\epsilon_-E^2}{d_-}+$$
$$\frac{90m^3\epsilon_+^2(m+E)^2}{d_+^2}-
\frac{15m^2\epsilon_+}{d_+}\big[m(7E^2+13m^2+20mE)+\epsilon_2(5m^2+6mE+M^2)\big]\,,$$

$$\overline{D}_{20}^1=\left| \overset{\rightharpoonup}{k_2}\right|  \left|
\overset{\rightharpoonup }{p_2}\right| +4 m^2 \Big(1-\frac{E}{d_-}\Big)-\frac{3 m \epsilon _+ (E+m)}{d_+}+\overset{\rightharpoonup }{k_2}{}^2
+\epsilon _2 (-E+m+\omega )\,,$$
$$D_{20}^{L_1}=-\omega ^2 [2 m (E+m)+M^2]+\omega  [5 m^3+2 E m \left(E+3 m-2 \epsilon _2\right)-2 M^2
   \epsilon _2]-m^3 \left(9 m+7 \epsilon _2\right)+$$
   $$M^2 \epsilon _2 \left(m-2\epsilon _2\right)+
   +2 E^2 m \left(\epsilon _2-2 m\right)-2 E m \left(3 m \left(2 m+\epsilon _2\right)+2 \epsilon _2^2\right)+\frac{E^2 m^2 \epsilon _-}{d_-}-$$
$$-\frac{6 m^2 \epsilon _+^2 (E+m)^2}{d_+^2}+\frac{m \epsilon _+}{d_+}[15 m^3+\epsilon _2 \left(7
   m^2+M^2\right)+7 E m (E+2 m)+2 E \epsilon _+ (E+5 m)]\,,$$

$$D_{20}^p = 4 m^2 E^2 \left(9 m^2+8 \epsilon _2 m+9 \epsilon
   _2^2\right)+m E [\left(63 m^2+17 M^2\right) m^2+4 \left(33 m^2+4 M^2\right) \epsilon _2 m+$$
   $$2 \epsilon _2^2 \left(51 m^2+10
   \epsilon _2 m+9 M^2\right)]-m^2M^4+m^4(18 m^2+29 M^2)+14 m^3 \epsilon _2\left(3 m^2+5 M^2\right)+$$
   $$+2m\epsilon _2^2 [\left(13 m+2 \epsilon _2\right) m^2+2 M^2 \left(15 m+4 \epsilon _2\right)]
   -Z\Big\{2 m [\left(11 m+29 \epsilon _2\right) m^2+\epsilon _2^2 \left(21 m+5 \epsilon
   _2\right)]+$$
$$+2 m E [8 m \epsilon _2+9 \left(m^2+\epsilon _2^2\right)]+\frac{m \epsilon _-^2\,M^2 (E-m) }{p^2}\Big\}+
\frac{\epsilon_+^2(m^2-M^2)}{m+E}\big[3\epsilon_+(m^2-M^2)+Z(2m+3\epsilon_2)\big]+$$
$$\frac{m^2(m^2-M^2)^2}{2Z}\Big[\frac{(m^2-M^2)^2}{Z}+M^2-m^2-4m\epsilon_2\Big]\,,$$

$$\overline{D}_{20}^p=-\frac{15m^5(m^2-M^2)^2}{2Z^2}+\frac{15m^5[m^2-2m(\omega-\epsilon_2)-3M^2]}{2Z}-\frac{15m^3\epsilon_+^2}{2d_+(m+E)}\big(18 m^2+30 E m+$$
$$4 \epsilon _2 m-M^2+13 E^2+4 E \epsilon _2\big)+15m^3[6m^2+7m\epsilon_2+2\epsilon_2^2+E(m+6\epsilon_2)]+10m^2\epsilon_-M^2-40m^2\epsilon_2E^2+ $$
$$5m\epsilon_-M^2E-20 d_-E^3(m+\omega)+2mp^2(4E^2+M^2)+5\omega^2(M^2E-2mE^2+8\epsilon_2E^2+6m^3-mM^2-$$
$$2\epsilon_2M^2)+\omega\big[15m^2(4\epsilon_2E-3mE-2E^2-2m\epsilon_+)+2p^2(4E^2+M^2)+5M^2(3m^2+\epsilon_2(m+E))+$$
$$10m\epsilon_2E^2\big]+10m^3E^2+\frac{45m^3\epsilon_+^3(m+E)}{d_+^2}-\frac{15m^3\epsilon_+^2}{2d_+(m+E)}\big(18m^2+30mE+4m\epsilon_2-M^2
+13E^2+4\epsilon_2E\big)-$$
$$\frac{15m^3\epsilon_-^2E}{2p^2d_-}(5E^2-4M^2-mE)\,,$$

$$\widetilde{D}_{20}=-\frac{15 \left(\left| \overset{\rightharpoonup }{k_2}\right|-\left|
   \overset{\rightharpoonup }{p_2}\right| \right){}^2\big(\omega^2-p^2\big)\big[2 E \left(m^2-\omega ^2\right)+E^2 (m+\omega )+\omega ^2 (\omega -m)\big]}{-2 d_- (E+m)+p^2+\omega  \left(\omega +2
   \epsilon _2\right)}+$$
$$-12(m+\omega ) \left(\left| \overset{\rightharpoonup }{k_2}\right|  \left|
   \overset{\rightharpoonup }{p_2}\right| +\epsilon _- \left(-E+\omega +\epsilon _2\right)\right){}^2+2\left(\left|
   \overset{\rightharpoonup }{k_2}\right|  \left| \overset{\rightharpoonup }{p_2}\right| +\epsilon _- \left(-E+\omega
   +\epsilon _2\right)\right)\big[10m^2(E-\epsilon_-)+$$
   $$10\omega(\omega+\epsilon_2)(\omega-2m)-10\omega^2(m+E)+(m+\omega)(10\epsilon_+E-4M^2-(E-\omega)^2)\big]-$$
$$-130 d_-
   m^2 \omega ^2+M^2 \big\{10 \big[-3 m^2 \left(-E+3 m+\epsilon _2\right)+E \omega  (m-2 \omega )-3 E \epsilon _2 (m+\omega )+4 m
   \omega  \epsilon _-+\omega ^2 \epsilon _2\big]+$$
$$+\omega ^2 (13 m-37 \omega )\big\}+2 p^2 \big\{10
   [E \left(m^2-\omega ^2\right)+m \epsilon _2 (\omega -m)]-25 [(m+\omega ) (m^2+m \left(\omega +\epsilon
   _2\right)+$$
   $$\epsilon _2 (E-\omega ))+\omega ^2 (\omega -2 m)]+\omega  (E+\omega ) (m+\omega )\big\}$$
$$+2 E \big\{10 m^2
   [\omega  \left(2 \left(m+5 \epsilon _2\right)+13 \omega \right)+3 m \epsilon _-]+\omega ^2 [\omega  \left(41 \omega
   +25 \epsilon _2\right)-m \left(89 \omega +35 \epsilon _2\right)]\big\}+$$
$$+\omega ^3 [\omega  (107 m-33 \omega )-50
   \left(m^2-2 m \epsilon _2+\omega  \epsilon _2\right)]+(m+\omega ) \left(17 E^2+8 M^2\right) \overset{\rightharpoonup
   }{p_1}{}^2\,,$$

$$D_{20}^1=
\left(\left| \overset{\rightharpoonup }{k_2}\right| \left| \overset{\rightharpoonup }{p_2}\right| +\epsilon _-
   \left(-E+\omega +\epsilon _2\right)\right) \Bigl(\frac{m^2 \epsilon _- p^2}{d_-}-\frac{2 m^2 \epsilon
   _+ (E+m)^2}{d_+}-\overset{\rightharpoonup }{k_2}{}^2 Z+m^4+$$
   $$+m^2 \left(\omega  (-E-m)+\epsilon _2 (m-\omega
   )\right)+m \epsilon _2 p^2+E m \left(\epsilon _+ (E+2 m)-2 \omega  \epsilon _2\right)-M^2 \omega
   \epsilon _2\Bigr)+$$
   $$+m \omega  \Biggl(-\frac{m \epsilon _- (3 E-m) p^2}{2 d_-}-\frac{m \epsilon _+ (E+m)
   \left(4 m (E+m)+p^2\right)}{2 d_+}-Z\overset{\rightharpoonup }{k_2}{}^2 +E
   \epsilon _2 \left(m^2+M^2\right)+$$
   $$+2 m^2 \left(m \epsilon _2-E^2\right)-\omega  \epsilon _- \left(E m+M^2\right)-E m \left(M^2+\epsilon
   _2^2\right)+m \epsilon _2 (4 E-\omega ) (E+m)\Biggr)\,.$$

\be
I_{11}=D_{11}+\frac{L_1}{k_h}D_{11}^{L_1}+\frac{L_2}{\omega}D_{11}^{L_2}+\frac{1}{\omega(2mp^2-d_-Z)}\Big[\frac{m-d_-}{k_h\,K_1}D_{11}^1+
\frac{p}{m}D_{11}^p\Big]\,,
\ee

$$D_{11}=-\frac{(m+\omega ) \left(\left| \overset{\rightharpoonup }{k_2}\right| +\left| \overset{\rightharpoonup
   }{p_2}\right| \right)}{m \omega }+\frac{m \epsilon _-}{2 d_-^2}+\frac{3 m-\epsilon _2}{4 d_-}-\frac{3 \epsilon _+}{4
   d_+}-\frac{\omega }{m}-1\,,$$

$$D_{11}^{L_1}=d_+-\frac{3 m^2 \epsilon _-}{2 d_-^2}+\frac{7 m^2-13 m \epsilon _2+4 \epsilon _2^2}{4
   d_-}-\frac{3 m \epsilon _+}{4 d_+}\,,$$

$$D_{11}^{L_2}=-\frac{m^2 \epsilon _-^2}{d_-^3}-\frac{\epsilon _-^2 \left(3 m-\epsilon _2\right)}{d_-^2}+\frac{\epsilon _- \left(5 m-3
   \epsilon _2\right)}{d_-}-2 \left(\omega -\epsilon _-\right)\,,$$

$$D_{11}^1=\left| \overset{\rightharpoonup }{k_2}\right|  \left|
   \overset{\rightharpoonup }{p_2}\right|  \Big[\frac{m^2 \epsilon _- p^2}{d_-^2}+\frac{m
   [\epsilon _2 p^2-\epsilon _- \left(E m+M^2\right)+]}{d_-}+\frac{m\epsilon_+(m+E)^2}{d_+}-\epsilon_2(mE+M^2)-$$
   $$-m\epsilon_+(m+E)\Big]+\frac{m \epsilon _+ (E+m)}{4 d_+}\big[-8 m^2 \epsilon _2+m
   M^2+E^2 \left(3 m-5 \epsilon _2\right)+4 E m \left(m-3 \epsilon _2\right)+M^2 \epsilon _2\big]+$$
$$\frac{m \epsilon _- }{4 d_-}\big[E \left(4 m^3-4 m^2 \epsilon _2+5 m M^2+3 M^2 \epsilon
   _2\right)-m M^2 \left(3 m+5 \epsilon _2\right)-E^3 \left(5 m+3 \epsilon _2\right)+E^2 m \left(7 m+\epsilon
   _2\right)\big]+$$
$$\frac{m^2 \epsilon _-^2(E+m) p^2}{2 d_-^2}+\omega\big[-Z\overset{\rightharpoonup }{k_2}{}^2 -m^2 \epsilon _2 (E+m)\big]+4E^2 m \overset{\rightharpoonup }{k_2}{}^2+m^3 M^2-E m^2 M^2+$$
$$+m^2 \big[2 \left(-E m^2+m^2 \epsilon _2+E^2 \epsilon _2\right)+\epsilon
   _2 \left(5 E m+M^2\right)\big]-\epsilon _2^3\,Z +$$
$$+\epsilon _2^2 [m^3+E \left(3 m^2+M^2\right)-mM^2]\,,$$
$$D_{11}^p=-\frac{m^3 \epsilon _-^2 (E-m)}{2 d_-^2}-\frac{3 m^2 \epsilon _+^2
   (E+m)}{4 d_+}+\frac{m^2 \epsilon _- [E \left(m+3 \epsilon _2\right)-m \left(3 m+\epsilon _2\right)]}{4 d_-}-$$
   $$-\omega ^2\,Z
   -\omega[m(m^2-2p^2)+\epsilon_2Z]+m[2m(E^2+m^2)+mE(3m-2\epsilon_2)-\epsilon_+M^2]\,.$$

\be
I_{10}=D_{10}-2mk_h\,L_1\,D_{10}^{L_1}+p\,D_{10}^p-\frac{1}{(2mp^2-d_-Z)}\Big[\frac{4m(m-d_-)}{\omega\,k_h\,K_1}D_{10}^1+\frac{p}{m+E}D_{10}^p\Big]\,,
\ee
$$D_{10}=-\frac{8}{3m}\Big[\frac{p_2(m+\omega)}{\omega}[d_-(m+E)-2p^2+\omega(\omega-\epsilon_2)]+\Big(\frac{m(5m^2-3m\epsilon_2+\epsilon_2^2)}{\omega}-4m^2
+\epsilon_2^2+$$
$$+3\epsilon_2(\omega-m)\Big)\Big(\left|
   \overset{\rightharpoonup }{k_2}\right| +\left| \overset{\rightharpoonup }{p_2}\right| \Big)\Big]+16m\epsilon_-+8\omega\epsilon_+
   -\frac{4 m^2
   \epsilon _-}{d_-}-\frac{12 m^2 \epsilon _+}{d_+}-\frac{8 \omega ^2 \left(m+\omega +3 \epsilon _2\right)}{3 m}\,,$$
 $$D_{10}^{L_1}=\frac{1}{m}\Biggl(\frac{m^2+8 m \epsilon _2-2 \epsilon _2^2}{d_+}+\frac{3 m^2-8 m \epsilon _2+2 \epsilon _2^2}{d_-}\Biggr)+m
   \left(-\frac{\epsilon _-}{d_-^2}-\frac{3 \epsilon _+}{d_+^2}\right)+4\,,$$

 $$D_{10}^1=2 \left| \overset{\rightharpoonup }{k_2}\right|  \left|
   \overset{\rightharpoonup }{p_2}\right|\Big[\frac{m^2 \epsilon _- p^2}{d_-}-\frac{2 m^2
   \epsilon _+ (E+m)^2}{d_+}-d_- (E+m) \left(m \epsilon _++E \epsilon _2\right)+d_+ \epsilon _2 p^2+$$
   $$ +m^2[Z+(\epsilon_2+M)(E+M)]\Big]+d_-[\overset{\rightharpoonup }{k_2}{}^2 \left(2 m^3+8 E m (E+m)+EM^2\right)
   +4 m^2 \epsilon _2 (E+m) \left(E+\epsilon _+\right)]+$$
   $$d_+ E M^2 \overset{\rightharpoonup }{k_2}{}^2+\frac{m^2 \epsilon
   _-^2 (E+m) p^2}{d_-}+\frac{m^2 \epsilon _+ (E+m) [4 (E+m) \left(2 m \epsilon _2+E \epsilon
   _-\right)+\epsilon _+ p^2]}{d_+}+$$
$$-2d_+d_-[\overset{\rightharpoonup }{k_2}{}^2Z+m^2\epsilon_2(m+E)]-2m[\overset{\rightharpoonup
   }{k_2}{}^2 \left(2 E^3+m M^2\right)+m \epsilon _2 (E+m) (p^2+2 m \epsilon _++4\epsilon_2E)]\,,$$

$$D_{10}^p=-2m^2\Big\{\frac{m^2-M^2}{E+m}\Big[\frac{1}{p^2}-\frac{2}{Z}+\frac{\omega}{2mp^2-d_-Z}\Big(\frac{Z}{p^2}-2\Big)\Big]
+\frac{\epsilon_-(E-m)}{d_-p^2}-\frac{3\epsilon_+}{d_+(E+m)}\Big\}+$$
$$\frac{8}{m}\Big(\omega(m+E)+\epsilon_2(E-m)-m^2-\frac{m^4}{Z}-\frac{2p^2}{3}\Big)-\frac{8}{3\omega}\big[2p^2-3(m^2+\epsilon_-E)\big]\,.$$

 \end{appendix}

%

\end{document}